\documentclass[a4paper,11pt]{article}
\pdfoutput=1
\usepackage{graphicx,rotating,hyperref,slashed,amsmath,xcolor,amssymb,amsfonts,colortbl,cite}
\makeatletter
\hypersetup{colorlinks,bookmarksopen,bookmarksnumbered,
linkcolor=blus,pdfstartview=FitH,urlcolor=rossos,citecolor=verde}
\allowdisplaybreaks	 
\numberwithin{equation}{section}

\hypersetup{%
    ,urlcolor=black 
    ,citecolor=black
    ,linkcolor=black
    }

\AtBeginDocument{
  \hypersetup{
    urlcolor=blue,
    citecolor=blue,
    linkcolor=blue,
  }%
}

\allowdisplaybreaks

\def\lsim{\mathrel{\rlap{\lower3pt\hbox{\hskip0pt$\sim$}}
   \raise1pt\hbox{$<$}}}         
\def\gsim{\mathrel{\rlap{\lower4pt\hbox{\hskip1pt$\sim$}}
   \raise1pt\hbox{$>$}}}         

 \newcommand{\sfootnote}[1]{} 
\definecolor{bluc}{cmyk}{1,1,0,0.1}
\definecolor{rossoCP3}{cmyk}{0,.88,.77,.40}
\definecolor{rosso}{cmyk}{0,1,1,0.4}
\definecolor{rossos}{cmyk}{0,1,1,0.55}
\definecolor{rossoc}{cmyk}{0,1,1,0.2}
\definecolor{verdes}{cmyk}{0.92,0,0.59,0.4}

\hypersetup{colorlinks, bookmarksopen, bookmarksnumbered,
citecolor=verdes, linkcolor=bluc, pdfstartview=FitH, urlcolor=rossos}

\newcommand{\mio}[1]{}

\definecolor{Gray}{gray}{0.95}

\usepackage{multicol}
\usepackage{color}
\definecolor{rosso}{cmyk}{0,1,1,0.4}
\definecolor{rossos}{cmyk}{0,1,1,0.55}
\definecolor{rossoc}{cmyk}{0,1,1,0.2}
\definecolor{blu}{cmyk}{1,1,0,0.3}
\definecolor{blus}{cmyk}{1,1,0,0.6}
\definecolor{bluc}{cmyk}{1,1,0,0.1}
\definecolor{verde}{cmyk}{0.92,0,0.59,0.25}
\definecolor{verdec}{cmyk}{0.92,0,0.59,0.15}
\definecolor{verdes}{cmyk}{0.92,0,0.59,0.4}

\def\circa#1{\,\raise.3ex\hbox{$#1$\kern-.75em\lower1ex\hbox{$\sim$}}\,}

\newcommand{\beq}{\begin{equation}}
\newcommand{\eeq}{\end{equation}}

\newcommand{\bea}{\begin{eqnarray}}
\newcommand{\eea}{\end{eqnarray}}
\newcommand{\be}{\begin{equation}}
\newcommand{\ee}{\end{equation}}
\newfam\rsfsfam
\def\mathscr#1{{\fam\rsfsfam\relax#1}}

\def\circa#1{\,\raise.3ex\hbox{$#1$\kern-.75em\lower1ex\hbox{$\sim$}}\,}
\makeatletter

\def\hhref#1{\href{http://arxiv.org/abs/#1}{arXiv:#1}} 

\newcommand{\doi}[1]{\href{http://dx.doi.org/#1}{[doi]}}

\def\hhref#1{\href{http://arxiv.org/abs/#1}{arXiv:#1}} 
 
\def\art{\@ifnextchar[{\eart}{\oart}}
\def\eart[#1]#2#3#4#5#6{{\rm #2}, {\em #3 \bf #4} {\rm (#6) #5} ({\em #1})}

\def\article{\@ifnextchar[{\earticle}{\oarticle}}
\def\oarticle#1#2#3#4#5#6{{\rm #1}, {\em ``#6''}, {\rm #2 #3 (#5) #4}}
\def\earticle[#1]#2#3#4#5#6#7{{\rm #2}, {\em ``#7''}, {\rm #3 #4 (#6) #5}  [\hhref{#1}]}
\def\hepart[#1]#2{{\rm #2, \em#1}}
\def\heparticle[#1]#2#3{#2, {\em ``#3''} [\hhref{#1}]}

\newcounter{alphaequation}[equation]
\def\thealphaequation{\theequation\hbox to
0.6em{\hfil\alph{alphaequation}\hfil}}
\def\eqnsystem#1{
\def\@eqnnum{{\rm (\thealphaequation)}}
\def\@@eqncr{\let\@tempa\relax \ifcase\@eqcnt \def\@tempa{& & &} \or
  \def\@tempa{& &}\or \def\@tempa{&}\fi\@tempa
  \if@eqnsw\@eqnnum\refstepcounter{alphaequation}\fi
\global\@eqnswtrue\global\@eqcnt=0\cr}
\refstepcounter{equation} \let\@currentlabel\theequation \def\@tempb{#1}
\ifx\@tempb\empty\else\label{#1}\fi
\refstepcounter{alphaequation}
\let\@currentlabel\thealphaequation
\global\@eqnswtrue\global\@eqcnt=0 \tabskip\@centering\let\\=\@eqncr
$$\halign to \displaywidth\bgroup \@eqnsel\hskip\@centering
$\displaystyle\tabskip\z@{##}$&\global\@eqcnt\@ne
\hskip2\arraycolsep\hfil${##}$\hfil& \global\@eqcnt\tw@\hskip2\arraycolsep
$\displaystyle\tabskip\z@{##}$\hfil
\tabskip\@centering&\llap{##}\tabskip\z@\cr}
\def\endeqnsystem{\@@eqncr\egroup$$\global\@ignoretrue} \makeatother

\definecolor{fiorentina}{rgb}{.5,0,.5}

\begin{document}

\setcounter{page}{1} \baselineskip=15.5pt \thispagestyle{empty}

\bigskip\

\vspace{1cm}
\begin{center}

{\fontsize{19}{28}\selectfont  \sffamily \bfseries {
 Ultracompact vector stars
}
}

\end{center}

\vspace{0.2cm}

\begin{center}
{\fontsize{13}{30}\selectfont  Gianmassimo Tasinato } 
\end{center}

\begin{center}

\vskip 8pt
\textsl{ Physics Department, Swansea University, SA28PP, United Kingdom }\\
\vskip 7pt

\end{center}

 \vspace{0.3cm}
\begin{abstract}
\noindent
We 
analytically investigate a new
 family of  horizonless compact objects in vector-tensor
 theories of gravity,  called ultracompact vector stars.  They are sourced by
  a vector condensate,  induced by a non-minimal coupling with gravity. They can be as  compact as
 black
 holes, thanks to their internal anisotropic stress. In the spherically symmetric case
   their   interior   resembles
   an isothermal sphere, with a singularity  that can be resolved by
    tuning   the available integration constants. The star interior   
     smoothly matches  to an  exterior Schwarzschild  geometry, with no need of extra energy momentum tensor   at the  star surface.  
 We analyse  the 
   behaviour of geodesics  within the star interior,  where  stable circular orbits are allowed, as well as trajectories crossing in both ways the  star surface. 
  We analytically study stationary    deformations of the vector field and of the geometry, which break spherical
  symmetry,  and whose features depend on the vector-tensor theory we consider. We  introduce and determine the vector magnetic  susceptibility   as a probe of the star properties, and we analyze how the rate of rotation of the star is affected by the 
  vector  charges.

\end{abstract}

\vspace{0.3cm}

\section{Introduction}

The new experimental opportunities offered by gravitational wave 
science motivate  the theoretical analysis of  compact objects in theories  beyond General Relativity (GR). 
   Encouraging us to take  a broader
  view on gravitational interactions, their study can    address, by means
  of concrete examples, open issues in our understanding of  gravity in strong-field regimes. For  instance: establish the validity of no-hair \cite{Bekenstein:1996pn} and cosmic censorship \cite{Wald:1997wa} hypothesis. Clarify the extremal properties of self-gravitating  compact objects \cite{Shapiro:1983du}, as a function of their content. Determine 
    the most  convenient methods
   for testing GR against alternative theories of gravity  \cite{Yunes:2016jcc}.
    See
e.g. \cite{Barack:2018yly} for a  review on these and related subjects.
Examples of exotic  compact objects relevant for our discussion are: boson stars \cite{Kaup:1968zz,Feinblum:1968nwc,Ruffini:1969qy},
first introduced as regular solutions of Einstein-Klein-Gordon equations (see e.g. \cite{Liebling:2012fv,Visinelli:2021uve} for  reviews), and
then shown to exist  also in the vector-tensor case  \cite{Brito:2015pxa,SalazarLandea:2016bys,Minamitsuji:2017pdr}. Gravastar, supported by negative pressure fluids \cite{Mazur:2001fv,Mazur:2015kia,Cattoen:2005he,Chirenti:2007mk} (see e.g. \cite{Cardoso:2019rvt} for a 
general discussion). The singular isothermal sphere  \cite{Tolman:1939jz,Oppenheimer:1939ne,Shu:1977uc,Christodoulou:1984mz,Remmen:2021tyj}, a self-similar configurations supported by  a perfect fluid in its interior  which contains a naked singularity \cite{Ori:1989ps}, and which  has been investigated also
in the context of 
       exceptions to the cosmic censorship hypothesis 
    (see e.g.
   \cite{Joshi:2011rlc} for a review). 
 
 \smallskip
 In this work we present and investigate  new analytic solutions
 describing a family of  compact objects in vector-tensor
 theories of gravity, which we dub ultracompact vector stars.  They correspond
 to a vector condensate,  induced by a non-minimal coupling with gravity \cite{Tasinato:2014eka,Heisenberg:2014rta,Gripaios:2004ms,Hull:2014bga,Allys:2015sht}
 that break the vector Abelian symmetry. The Lagrangian for the system has no additional parameters with respect to an Einstein-Maxwell system. 
  The vector field  can be interpreted as a dark photon with relevant applications
  for dark energy (see e.g. \cite{Tasinato:2014mia,DeFelice:2016yws}), but it can also be motivated by  recent theoretical advances in characterising  scenarios of ultralight
  vector dark matter \cite{Graham:2015rva,Agrawal:2018vin,Dror:2018pdh,Co:2018lka,Bastero-Gil:2018uel} (see e.g. \cite{Antypas:2022asj} for a recent review).   We refer also to \cite{Fabbrichesi:2020wbt,Caputo:2021eaa} for  recent
  reviews on the physics of dark photons, and their phenomenological consequences. 

 The self-gravitating objects can be as compact as black
 holes, thanks to their internal anisotropic stress: we discuss their properties
 in section \ref{sec_sphersyms}. In fact, a free
 parameter controls their compactness,   spanning from Minkowski space
 to the black hole solutions first studied in  \cite{Chagoya:2016aar} (see \cite{Chagoya:2017fyl,Minamitsuji:2016ydr,Babichev:2017rti,Heisenberg:2017xda,Heisenberg:2017hwb,Filippini:2017kov,Kase:2017egk,Chagoya:2017ojn,Kase:2018owh} for further developments).
  In the spherically symmetric case, their
 external geometry corresponds to the Schwarzschild solution as
 in GR, even if they are characterised by an electric-type charge. The solution has no horizon, since its Schwarzschild  radius
 is located inside the star.  The interior part of the solution resembles
   a  singular isothermal sphere, and the singularity at the star centre can be resolved by
    tuning some of the available integration constants. The interior configuration  
    is smoothly matched to the exterior geometry of the star, with no need of extra energy momentum tensor localised at the star surface.  
 
 By studying the behaviour  of geodesics and of vector and metric perturbations, we investigate
 how to distinguish vector star solutions from self-gravitating configurations in GR. In fact, the  simplicity of our solutions allow us to analytically study 
  in detail their properties. 
  The study of geodesics in section \ref{sec_geodesics} reveals new features with respect to Schwarzschild black 
 holes in GR, thanks to the possibility to cross the star surface in both directions. We find new stable circular orbits for time-like geodesics whose location depend on the star compactness. We also  study geodesic trajectories that enter, bounce, and leave from  the star interior, with the  time spent to complete the process depending on the star properties.  For  null-like geodesics, we relate as \cite{Cardoso:2017cqb} the presence of unstable circular orbits (light-rings) with 
 parameters controlling the  compactness of the object (see also \cite{Cardoso:2019rvt} for 
 a comprehensive review on how to distinguish exotic compact 
 objects from black hole configurations). 
 
  In section \ref{sec_beyond}
 we analyse stationary, parity-odd fluctuations of the system. The inclusion
 of magnetic vector perturbations allows us to switch on a magnetic-type  charge, 
which  backreacts on the geometry
at the linearised level, and breaks the   spherical symmetry  forcing the star to slowly rotate.  The perturbed geometry  
 depends on the features of the vector profile, making the exterior configuration distinguishable
 from their GR counterparts.  
The  rotation rate depends both on the values of magnetic and electric charge, and    the interior region of the object
is dragged by the external rotation. We  study the response of the vector field
profile to an external magnetic field, and the distinctive features of the induced  magnetic susceptibility.

 We conclude in section \ref{sec_outlook}, discussing possible future directions for studying the properties
of vector star configurations.


\section{Spherically symmetric  vector stars}
\label{sec_sphersyms}
 
 \subsection{The set-up and the field equations}
We consider a vector-tensor theory described by the following Lagrangian density (we set $M_{\rm Pl}\,=\,c\,=\,\hbar\,=\,1$)
\be \label{lagint1}
 {\cal L}\,=\,\frac{R}{2}-\frac14\,F_{\mu\nu}^2+\frac{\beta}{4}\,V_\mu \,V_\nu\,G^{\mu\nu}
 \,,
 \ee
 with $V_\mu$ the vector field and $F_{\mu\nu}\,=\,\partial_\mu V_\nu-\partial_\nu V_\mu$
 the associated field strength. 
The last term, proportional to the dimensionless parameter $\beta$, controls a ghost-free \cite{Tasinato:2014eka,Heisenberg:2014rta,Gripaios:2004ms} non-minimal coupling of vector fields with the Einstein
tensor $G_{\mu\nu}$. This coupling with gravity  breaks the   Abelian gauge symmetry $V_\mu\to V_\mu-\partial_\mu \chi$, with $\chi$  an arbitrary scalar field. Notice
that we do not include a mass term for the vector:  the gauge symmetry breaking
is only due to coupling with gravity. It would be interesting
to find symmetry arguments protecting the structure of the theory governed
by Lagrangian \eqref{lagint1}, for example
along the lines of \cite{Tasinato:2020fni}. While Lagrangians as  \eqref{lagint1} have
been studied at length in the context of dark energy, it would
 be interesting to explore possible applications
for dark matter, as  for the ultralight vector dark matter scenarios of \cite{Graham:2015rva,Agrawal:2018vin,Dror:2018pdh,Co:2018lka,Bastero-Gil:2018uel}.

\smallskip
  From now on,  we make the choice
\be
\beta=1\,,
\ee
since it is the simplest option for finding novel  configurations
with interesting properties. This choice implies  that we do not have
additional free parameters with respect to the standard Einstein-Maxwell system. 
 In this section
  we focus on spherically symmetric solutions associated with 
 the metric element
 \bea
 \label{metrans1}
 d s^2\,=\,g_{\mu\nu}\,d x^\mu d x^\nu&=&-A(r)\,d t^2+\frac{d r^2}{B(r)}\,+r^2\, d\theta^2+r^2\,\sin^2{\theta}\, d\phi^2\,,
\eea
and an electric-type vector field Ansatz
\bea
 \label{vecans1}
V_\mu d x^\mu
&=&\alpha_0(r)\,d t+\Pi(r)\,d r\,,
 \eea
 associated with a dark electric charge for the object. 
  In section \ref{sec_beyond} we  extend our discussion to 
  configurations with magnetic-type charges, and solutions  breaking spherical symmetry.  
 Notice that
 the metric can be influenced by the vector radial
   vector profile  $\Pi(r)$,  since the theory we
  consider breaks the Abelian gauge symmetry and the radial  vector component
  is not removable by a gauge transformation.
   The field equations  associated with our  Ans\"atze of eqs \eqref{metrans1} and \eqref{vecans1} are  (all quantities depend on the radial direction $r$ only)
\bea
0&=&\Pi\,\left( A'-\frac{A}{r B} \left(1-B\right)\right) \label{eqfPI}
\,,
\\
0&=&\hskip-0.3cm\frac{\alpha_0'^2+\partial_r \alpha_0^2 }{4 A}+\frac{\alpha_0^2(B-1)}{8 r^2\,A\,B}
+\left( \frac{4+3 B \Pi^2}{8\,r\,A}-\frac{\alpha_0^2}{8\,r\,A^2}\right)\,A'
+\frac{B\left( 4+(3 B-1)\,\Pi^2\right)-4}{8\,r^2\,B}
\,,
\label{eqffB}
\\
0&=&\partial_r\,\left[\frac{r^2\,B^{1/2}\,\alpha_0'}{A^{1/2}}-\frac{r\,\alpha_0}{A^{1/2}\,B^{1/2}} \right]
+B^{-1/2}\,\partial_r \left( \frac{r\,\alpha_0}{A^{1/2}}\right)
+\frac{\alpha_0\,(B-1)}{2\,A^{1/2}\,B^{1/2}}
\,,
\eea
and
\bea
0&=&\partial_r \left[ B^{3/2}\,\Pi^2+B^{1/2} \left(1-\frac{\alpha_0^2}{4 \,A} \right) \right]+B^{1/2}\,\partial_r \left( \frac{\alpha_0^2}{4 A }
\right)+\frac{r\,B^{1/2} \alpha_0'^2 }{A}\nonumber
\\
&&+\frac{B^{1/2}\,\left(B-1\right)\,\Pi^2}{2 r}+\frac{\alpha_0^2\,(B-1)}{2 r\,B^{1/2}\,A}+
\frac{2 B-2}{r\,B^{1/2}}\,.
\label{eqfB1}
\eea
The condition \eqref{eqfPI} identifies two branches of solutions. 
 If we were considering $\beta$ a free parameter, 
the branch with $\Pi(r)\,=\,0$ would be  continuously connected with the Reissner-Nordstr\"om configuration, when  sending $\beta$ to zero  \cite{Chagoya:2016aar}. We concentrate here on the second branch with $\Pi(r)\,\neq\,0$, that exists
because of the non-minimal coupling with gravity in the Lagrangian \eqref{lagint1}. Eqs \eqref{eqfPI} and \eqref{eqffB} are algebraic equations and  dictate the conditions
\bea
\label{condfB1}
B(r)&=&\frac{A(r)}{A(r)+r\,A'(r)}
\,,
\\
\label{condfPI1}
\Pi^2(r)&=&\frac{r\,\alpha_0^2(r)}{A(r)}\,\left(\frac{A'(r)}{A(r)}-2\frac{\alpha_0'(r)}{\alpha_0(r)}-r\,\frac{\alpha_0'^2(r)}{\alpha_0^2(r)} \right)
\,.
\eea
The Ricci scalar  associated with the metric \eqref{metrans1}, and with the condition \eqref{condfB1}, is
\be
\label{eqRicci1}
{\text{Ricci scalar}}\,=\,
\frac{\left(2 A(r)-r \,A'(r) \right) \left(  2 A'(r)+r A''(r)\right)}{2 r\,\left( A(r)+r\,A'(r)\right)}\,,
\ee
and usually diverges at the origin $r=0$, unless $A(r)$ acquires a specific profile (more on this later). 
 Once we 
implement   the
 conditions \eqref{condfB1} and \eqref{condfPI1}, it is straightforward to check 
that a solution of eqs \eqref{eqfPI}--\eqref{eqfB1} is~\footnote{A version of these solutions was presented in \cite{Minamitsuji:2016ydr}. See also the discussion in \cite{Babichev:2017rti}.}
\bea
\label{gensol1}
\alpha_0(r)&=&\frac{2\,Q\,R}{r}+2 \sigma+\frac{2\,\left(1-\sigma\right)}{1+\gamma}\,\left(\frac{r}{R}\right)^{\gamma}\,,
\\
\label{gensol2}
A(r)&=&\sigma^2-\frac{2 M}{r}+\frac{2\,\sigma\,\left(1-\sigma\right)}{(1+\gamma)}\,\left(\frac{r}{R}\right)^\gamma+\frac{\left(1-\sigma \right)^2}{(1+2\gamma)}\,\left(\frac{r}{R}\right)^{2\gamma}\,,
\\
B(r)&=&\frac{A(r)}{\left( \sigma+\left( 1-\sigma\right) (r/R)^\gamma\right)^2 }
\label{gensol3}\,,
\eea
where $Q$,  $\gamma$ and $\sigma$ are dimensionless constant parameters, while  $M$, $R$ dimensionful parameters. The solution for $\Pi(r)$ is cumbersome and we
do not write it explicitly: it is directly  obtained plugging eqs \eqref{gensol1} and \eqref{gensol2} in eq \eqref{condfPI1}.
Choosing $\gamma=0$, $\sigma=1$ one finds the black hole solutions of \cite{Chagoya:2016aar}: the geometry corresponds to a stealth
Schwarzschild space-time, with a non-vanishing profile for $\alpha_0(r)$ and $\Pi(r)$ and a horizon at $r\,=\,2 M$.  
Turning on $\gamma$ and $\sigma-1$ allows us to determine  new
families  of  horizonless configurations: these parameters have a transparent physical interpretation, that we are going to discuss in what
comes next.

 \subsection{The simplest ultracompact vector star configuration}
 \label{sec_simplest}
 
 We   proceed asking:
   {Can we use the solutions  \eqref{gensol1}--\eqref{gensol3}  as building blocks to construct     horizonless   objects, with  compactness 
  ${\cal C}$ in the interval $0\le {\cal C}<1/2$, and with   an  
  asymptotically flat exterior geometry? }

\smallskip

We  are going to answer affirmatively to the question, by discussing a simple analytical configuration with the desired properties. 
 The horizonless object  can be as compact as a Schwarzschild black hole,  violating the Buchdahl bound \cite{Buchdahl:1959zz}  thanks to the anisotropic  stress characterising its   interior.
 For this reason we call the solution ultracompact vector star. (See e.g. \cite{Raposo:2018rjn} and references therein for a recent analysis of consequences of 
 compact objects with pronounced anisotropic stress tensor.)
 The star interior geometry is sourced by a vector condensate, induced
 by the non-minimal couplings with gravity in Lagrangian \eqref{lagint1}.
 As we will learn, the simplicity 
of the configuration will allow us to analytically investigate many of  its properties.

\subsubsection*{The  geometry of the vector star}

 We assume that the  geometry is separated in two regions,  smoothly matching  at a radial position $R$, which we interpret as    the radius  of the compact object. In fact, 
  we aim at finding a configuration with a continuous transition
from an  interior to an exterior region at   $r\,=\,R$, with no
need of  additional localised energy-momentum tensor at the star surface $r=R$.

\smallskip

 For characterizing  the {\bf interior region} $r\le R$ we take the configuration
\eqref{gensol1}, \eqref{gensol2}, and set $M\,=\sigma\,=\,0$ focussing on $\gamma\ge0$. The line element $d s^2\,=\,g_{\mu\nu} d x^\mu d x^\nu$, with  $g_{\mu\nu}\,=\,{\rm diag}(-A(r),\,1/B(r),\,r^2,\, r^2 \sin^2{\theta})$, results
 \bea
 \label{geoss1}
 d s_{\rm (int)}^2&=&-\frac{1}{(1+2\gamma)}\,\left(\frac{r}{R}\right)^{2\gamma}\,d t^2+(1+2\gamma)\,{d r^2}\,+r^2\, d\theta^2+r^2\,\sin^2{\theta}\, d\phi^2\,.
\eea
The coefficients of the vector field configuration,
$
V_\mu d x^\mu
\,=\,\alpha_{\rm (int)}(r)\,d t+\Pi_{\rm (int)}(r)\,d r
$, 
are  
\bea
\alpha_{\rm (int)}(r)&=&\frac{2 Q_I\,R}{r}+\frac{2}{1+\gamma}\,\left(\frac{r}{R}\right)^{\gamma}\,,
\label{scalALintsol}
\\
\Pi^2_{\rm (int)}(r)
&=&4\, \left(1+2\gamma\right)^2\,\left[
\left(\frac{Q_I\,R^{1+\gamma}}{r^{1+\gamma}}+\frac{1}{1+\gamma} \right)^2-
   \frac{1}{1+2\gamma }
\right]\,,
\label{scalPintsol}
\eea
where $Q_I$ the vector charge in the interior geometry.
   The solution \eqref{geoss1}-\eqref{scalPintsol} is valid for any radius $r \le R$: for
ensuring that the profile for $\Pi_{\rm (int)}(r)$ is real in this interval, we impose the following inequality on the charge $Q_I$:
\be
\label{conQI}
Q_I\,\ge\,\left( \frac{1}{\sqrt{1+2 \gamma}}-\frac{1}{1+\gamma}\right)\,.
\ee
The function in the right hand side of eq \eqref{conQI} is positive and has a maximum for $\gamma\simeq
5.22$, with value $0.13$: a $Q_I\,>\,0.14$ then ensures that \eqref{conQI}
is satisfied for any value of $\gamma$. 
The geometry described by  eq \eqref{geoss1} is quite simple, and corresponds to a self-similar space-time   controlled by the constant parameter $\gamma$. In fact,
 the parameters have been chosen in such a way to relate our geometry to the one
 of the singular isothermal sphere \cite{Tolman:1939jz,Oppenheimer:1939ne,Shu:1977uc,Christodoulou:1984mz}. 
  Under a scale transformation $r \to \lambda\,r$, $t\to \lambda^{1-\gamma}\,t$, the metric scales as 
 $d s^2\,\to\, \lambda^2\,d s^2$.
  Examples of self-similar geometries have been found in the literature \cite{Ori:1989ps,Remmen:2021tyj},  sourced by a perfect fluid equation 
of state. 

 In our case
  the interior geometry is sourced by  vector field condensate, induced by
   the non-minimal couplings of the vector with gravity: the corresponding energy-momentum tensor associated
  with the vector field \footnote{Notice that we could also include
  further sources of internal stress tensor, for example an additional perfect fluid: the  solutions
  can be  extended to include this case as well.} can be computed straightforwardly using our Ansatz for the interior star configuration. It is anisotropic and can be expressed as
  \bea
  T_\mu^{\,\nu}\,=\,{\rm diag} \left(-\rho,\,p_r,\,p_t,\,p_t\right)\,,
  \eea  
  with
  \bea
  \rho_{r}&=&\frac{A(r) \left( 2 A'(r)+r A''(r)\right)}{r\,\left( A(r)+r A'(r) \right)^2}\,=\,\frac{2 \gamma}{(1+2\gamma)\,r^2}
  \,,
\\
p_r
&=&0\,,
\\
p_t&=& \frac{r\,A'(r)}{4\,A(r)}\,
\rho_r\,=\,\frac{ \gamma^2}{(1+2\gamma)\,r^2}\,=\,\frac{\gamma}{2}\,\rho_r \,.
  \eea
 While the first of the equalities in the previous equations applies to any solution
 of the system  \eqref{eqfPI}--\eqref{eqfB1} (selecting the branch with
 non-vanishing profile $\Pi(r)$), the remaining ones are  specialized
 to the internal configuration  \eqref{geoss1}. Notice that the tangential pressure is larger than 
 the radial one, $p_t\ge p_r$, hence we have the opportunity to overcome
 Buchdahl theorem  \cite{Buchdahl:1959zz,Cardoso:2019rvt}.
 
   The interior geometry \eqref{geoss1} is singular, since it has
a singularity at the origin $r=0$ as it can be readily checked computing the Ricci scalar.   (Self-similar configurations are being discussed
 as possible candidates for the exceptions to the cosmic censorship hypothesis,
 see  e.g.
   \cite{Joshi:2011rlc} for a review.) 
   In our system, the singularity can be resolved turning
on the parameter $\sigma$ in eqs \eqref{gensol1} and \eqref{gensol2}:  we will study this 
option in   section \ref{sec_resolving}.  We find this feature interesting, since  the singularity can be resolved
without changing the initial action.

\smallskip

For characterizing the   {\bf exterior region} $r\ge R$ we instead set $\gamma=0$, $\sigma=1$ in eqs \eqref{gensol1} and \eqref{gensol2}. The resulting line element 
corresponds to a Schwarzschild geometry
 \bea
 \label{geoss2}
 d s_{\rm (ext)}^2&=&-\left(1-\frac{2 M}{r}\right)\,d t^2+\frac{d r^2}{\left(1-\frac{2 M}{r}\right)}\,+r^2\, d\theta^2+r^2\,\sin^2{\theta}\, d\phi^2\,.
\eea
 The vector field profile $
V_\mu d x^\mu
\,=\,\alpha_{\rm (ext)}(r)\,d t+\Pi_{\rm (ext)}(r)\,d r
$ is
\bea
\label{solVext1}
\alpha_{\rm (ext)}&=&2+\frac{2\,Q_{E}\,R}{r}
\,,
\\
\Pi^2_{\rm (ext)}&=&\frac{4\,Q_{E}^2\,R^2+8 M \,r+8 Q_{E}\,R\,r}{(r-2M)^2}
\,.
\eea
The quantity $Q_E$ corresponds to an electric-type charge for the star configuration,
as measured by an external observer.  
In order to smoothly {match} the interior and exterior geometries at radius $R$, we select the exterior mass $M$ and the exterior charge $Q_{E}$ as
\bea
\label{relQEI}
Q_{E}&=&{Q}_I-\frac{ \gamma}{1+\gamma}\,,
\\
\label{relMG1}
M&=&\frac{\gamma}{1+2 \gamma}\,R\,,
\eea
where $Q_I$ is the vector charge in the interior, as appearing in eq \eqref{scalALintsol}. 
Interestingly, 
these conditions {\bf automatically ensure} that the profiles for $A$, $B$, $\alpha_0$, $\Pi$
are continuous at $r=R$, as well  as~\footnote{ Notice that it is not necessary to impose that
the first derivatives of $B$ and $\Pi$ are continuous
at the surface $r=R$, since these  components on the radial direction along which we match the geometries. This can 
be checked by computing the Israel junction conditions. Alternatively, we can 
 integrate  equations \eqref{eqfPI}-\eqref{eqfB1} along the radial direction, over a small interval $R-\epsilon$, $R+\epsilon$ for  an infinitesimal $\epsilon$. One can check that the consistency of the equations requires the continuity of all functions at $r=R$, but only  of the first derivatives of $A$, $\alpha_0$
 at that position.}   the first derivatives of $A$, $\alpha_0$, with no need of extra energy momentum tensor at the matching surface position.  This differs from gravastar configurations,
 which make use of a crust of energy momentum tensor for connecting the exterior
 with the interior geometry \cite{Mazur:2001fv}. Using this property, in what follows we
 {\bf define} the star surface as the position $r=R$ where the star interior
 and exterior geometries smoothly match. 
 
 \smallskip

\begin{figure}
\begin{center}
  \includegraphics[width = 0.45 \textwidth]{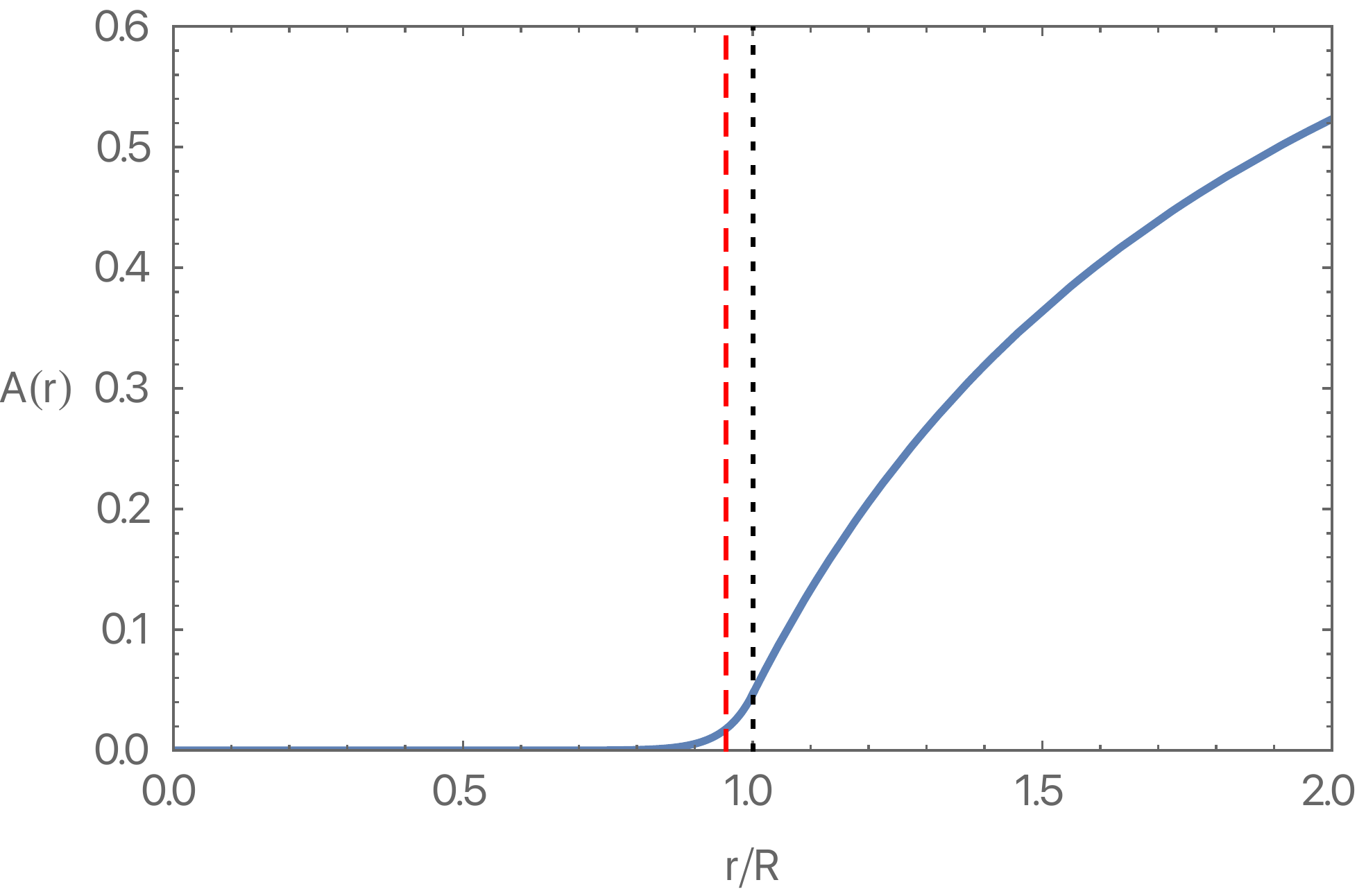}
    \includegraphics[width = 0.45 \textwidth]{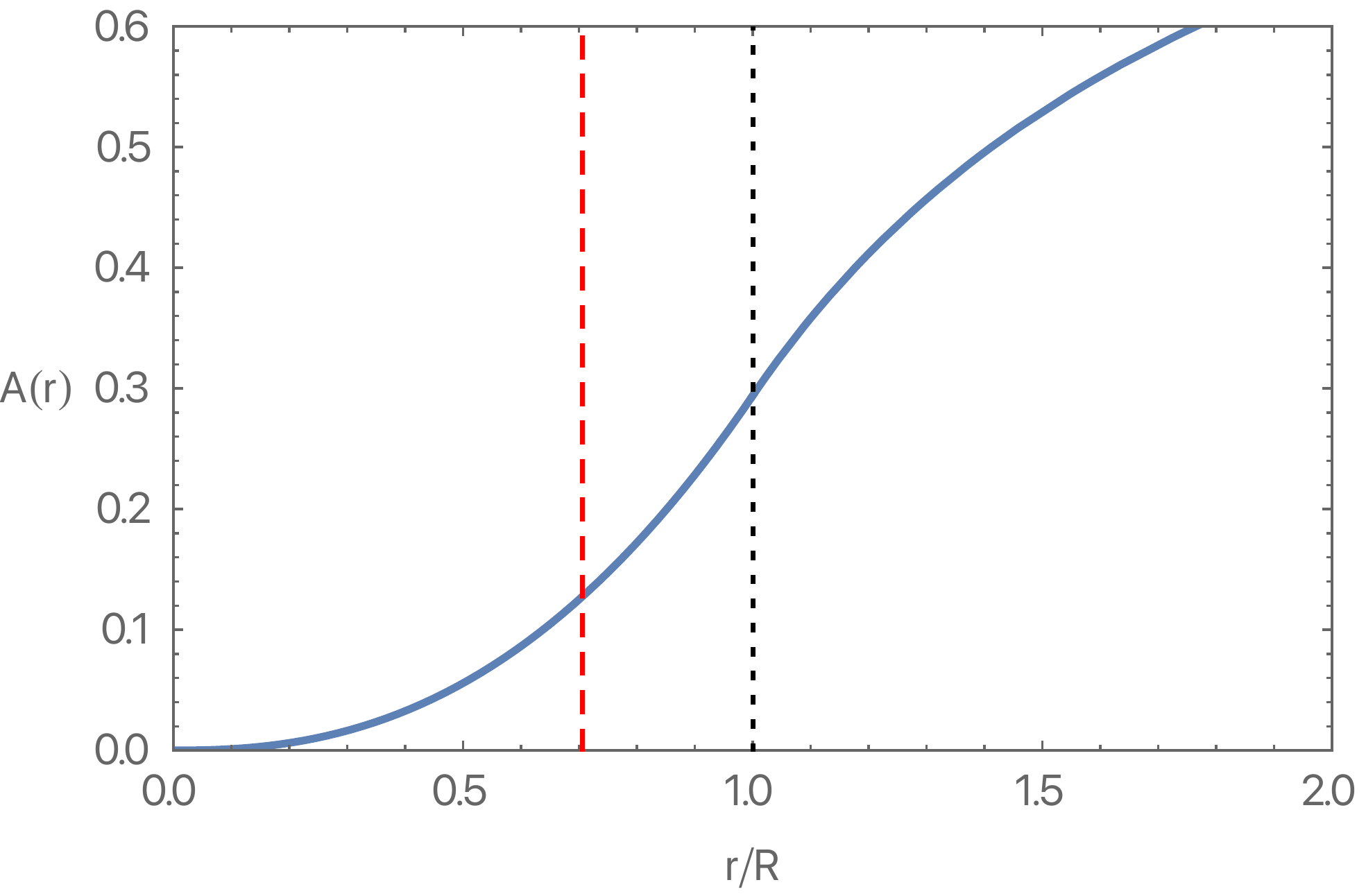}
  \includegraphics[width = 0.45 \textwidth]{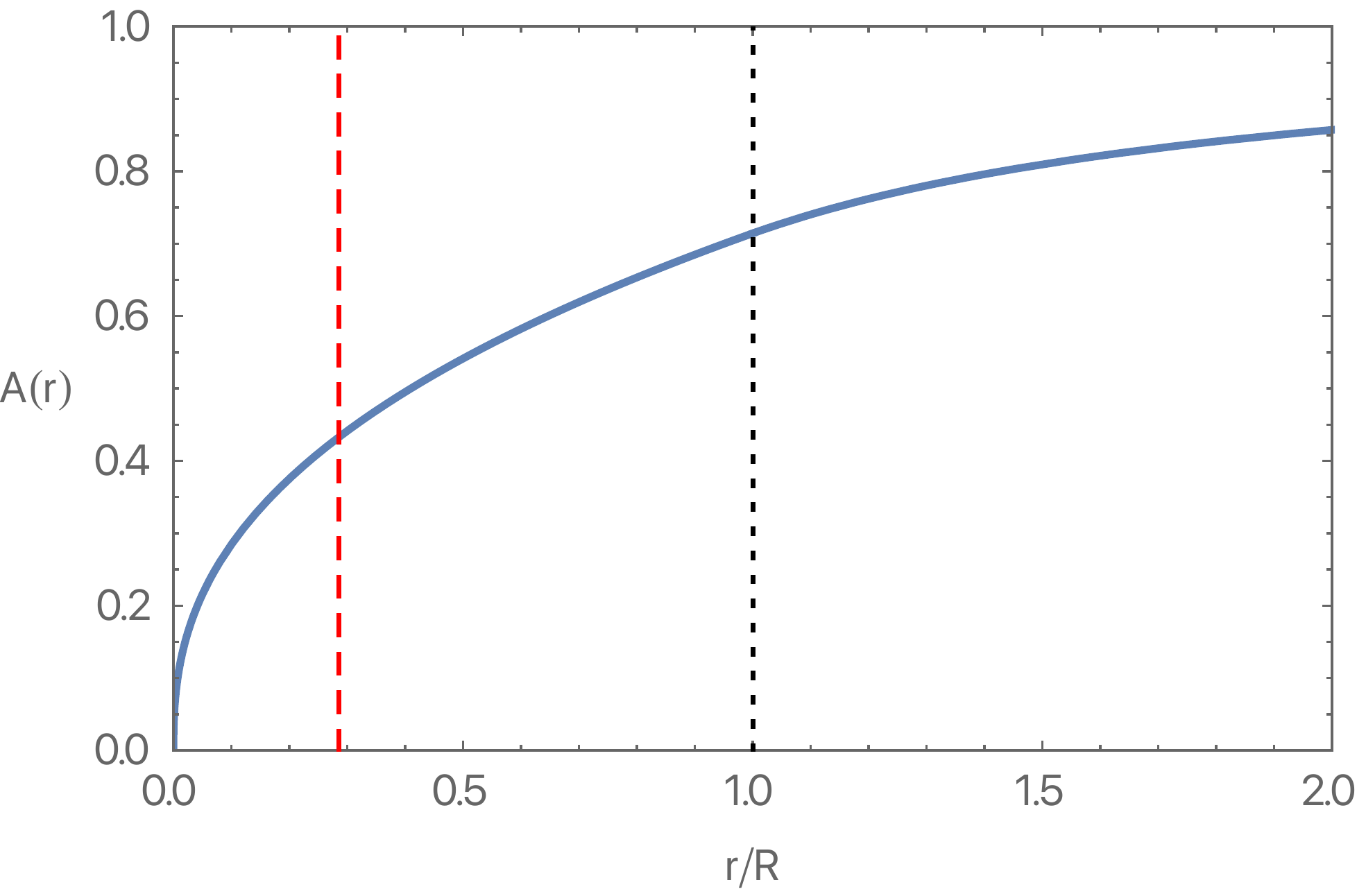}
    \includegraphics[width = 0.45 \textwidth]{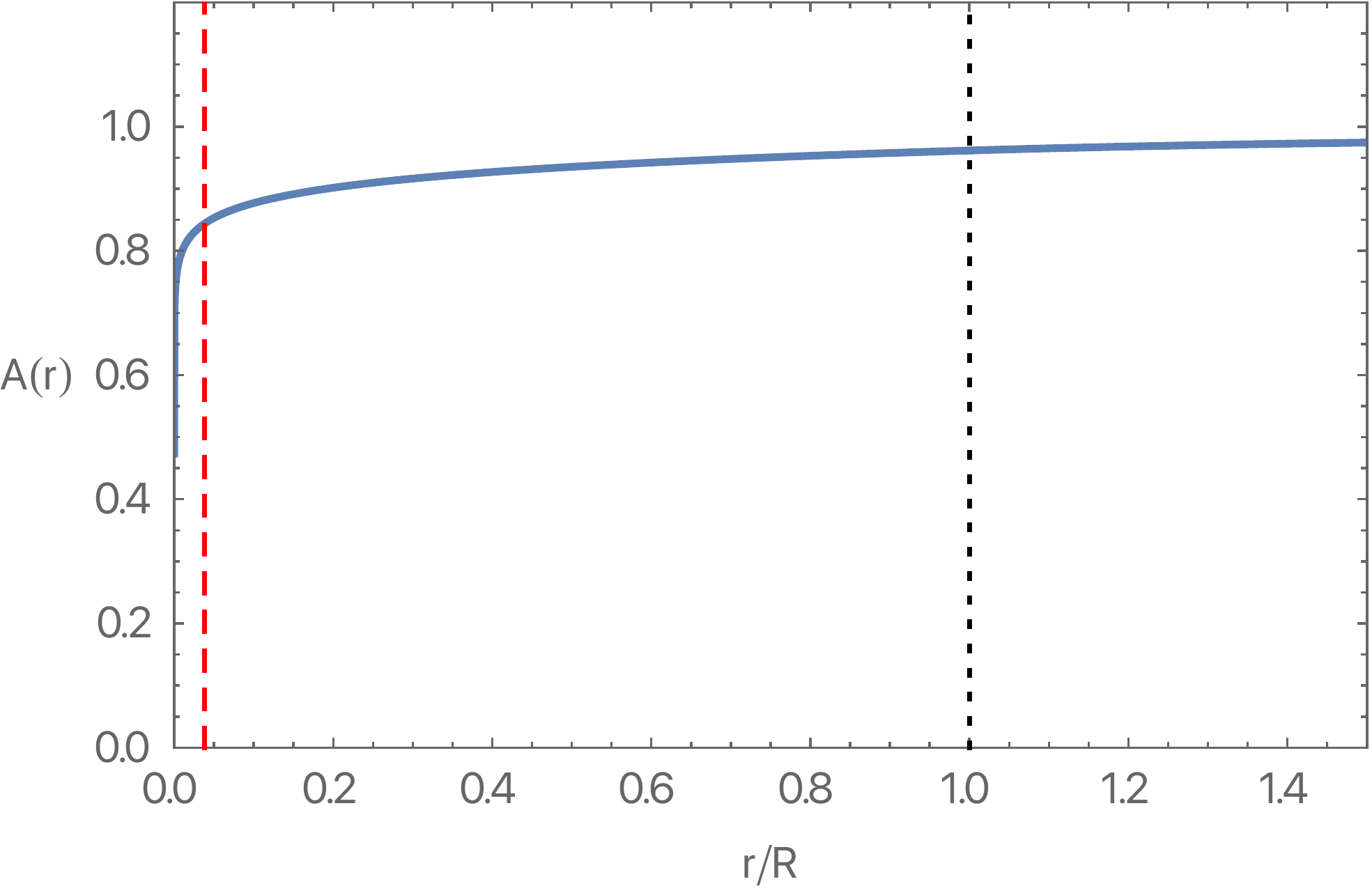}
    \end{center}
 \caption{\it \small The time-time metric component $A(r)$ (see our metric Ansatz  in eq \eqref{metrans1}) both in the interior and exterior of the star. We make 
   the following choices (from left-to-right and top-to-bottom) of the exponent $\gamma$: $\gamma=10.2, \,1.2,\,0.2,\,0.02$. Red dashed line: the Schwarzschild radius $r=2 M$. Black dashed line: the position of the star surface $r=R$. }
  \label{fig_confinfA}
\end{figure}

 We turn off  possible additional contributions associated
 with parameters $\sigma-1$, $\gamma$ in eq \eqref{gensol2} in the exterior
 region of the  geometry, since
 we find that only the   Schwarzschild profile in eq \eqref{geoss2} allows us to consistently satisfy the matching conditions at the sphere radius $R$. We interpret this fact  as  a manifestation of a no-hair condition for this set-up, so that in the spherically symmetric case the exterior
 geometry is only characterised by the  mass $M$, while it is independent from the charge  $Q_E$. We will learn in section \ref{sec_beyond} 
    that the story is more complex (and interesting) when breaking
 spherical symmetry, since then the geometry  depends  on magnetic and
 electric charges.
 
\subsubsection*{The vector star compactness}

We find the inequality
\be
r_{\rm Schw}\equiv 2 M\,=\,\frac{2 \,\gamma\,R}{1+2 \,\gamma}\,\le\,R\,,
\ee
showing that the Schwarzschild radius $r_{\rm Schw}$ is inside the star radius $R$. We conclude that our  configuration
has no horizons. Hence the non-negative parameter $\gamma$ has a transparent physical interpretation. It controls the
{\it compactness} 
\be
{\cal C}\,\equiv\, M/R
\ee
 of the object we are examining. We find
\be
\label{defCom1}
0\,\le\, {\cal C}\,=\,\frac{ \gamma}{1+2 \gamma}\,<\,\frac12\,.
\ee
Hence
this quantity  spans from $0$ for $\gamma=0$ (flat space) to $1/2$ for $\gamma=\infty$, which is the same compactness of a Schwarzschild black
hole. As anticipated above, the vector star  can violate the Buchdahl bound ${\cal C}\le4/9$, thanks to a 
sizeable anisotropic stress.

We represent in Fig \ref{fig_confinfA} the time-time  metric component  $A(r)$,
  together with the position of the would-be Schwarzschild radius, for different choices of $\gamma$.  As  $\gamma$ becomes smaller and smaller,  the geometry approaches flat space.  In the opposite case, the larger the $\gamma$ is, 
the nearer the Schwarzschild radius approaches from inside the surface of the star.
 In the limit $\gamma\to\infty$ the interior geometry is not well defined, at least in Boyer-Lindquist coordinates, since the metric profile for $A$ becomes flatter and flatter (see Fig \ref{fig_confinfA}, upper left panel). However, as will discuss in what comes next,  this limit can be meaningful in
 specific contexts and applications. 

 \subsection{Resolving the singularity in the   star interior}
 \label{sec_resolving}

The singularity at the star origin $r\,=\,0$ can be resolved by switching on the parameter
$\sigma$ within the interval 
$0\le\sigma\le1$ in eqs \eqref{gensol2}, at least for large enough values of $\gamma$. 
 In order to obtain a configuration
   with a smooth matching at the star surface, 
 the interior geometry  and vector field solutions  are  given by (the cumbersome
  expression for $\Pi(r)$ can be determined plugging the next formulas in eq \eqref{condfPI1}):
\bea
\label{gensol1a}
\alpha_0(r)&=&\frac{2\,Q_I\,R}{r}+2 \sigma+\frac{2\,\left(1-\sigma\right)}{1+\gamma}\,\left(\frac{r}{R}\right)^{\gamma}
\,,
\\
\label{gensol2a}
A(r)&=&\sigma^2+\frac{2\,\sigma\,\left(1-\sigma\right)}{(1+\gamma)}\,\left(\frac{r}{R}\right)^\gamma+\frac{\left(1-\sigma \right)^2}{(1+2\gamma)}\,\left(\frac{r}{R}\right)^{2\gamma}
\,,
\\
B(r)&=&\frac{A(r)
}{\left[
 \sigma+\left( 1-\sigma\right) (r/R)^\gamma\right]^2}
\label{gensol3a}
\,.
\eea
The exterior geometry is again given by the Schwarzschild expressions \eqref{geoss2}, but this time the mass
and the outside vector charge are given by
\bea
M&=&\frac{(1-\sigma)\,\gamma\,(1+\gamma+\sigma\,\gamma)}{(1+\gamma) (1+2\gamma)}\,R
\,,
\\
Q_{\rm E}&=&{Q}_I-\frac{ \gamma(1-\sigma)}{1+\gamma}
\,,
\eea
in order to continuously match the interior with the exterior. 
This configuration is typically less compact than the one of section \ref{sec_simplest}, since now the
  compactness parameter ${\cal C}$ spans between
 ${\cal C}=0$ for $\gamma=0$, to ${\cal C}\,=\,(1-\sigma^2)/2$ for $\gamma\to\infty$ (recall
 that $0\le\sigma\le1$).  

The curvature invariants are now no more necessarily singular for $r\to 0$. In fact, computing for example the Ricci scalar by means of eq \eqref{eqRicci1} we find
\bea
\hskip-0.6cm
{\text{Ricci scalar}}\,=\, \frac{2\,r^{\gamma-2}\,\gamma\,(1-\sigma)}{\left[\sigma+(1-\sigma)\,r^{\gamma} \right]^3}\,\left[\sigma^2+\frac{\sigma\,(2-\gamma)\, (1-\sigma)}{1+\gamma} 
\,r^\gamma+\frac{(1-\gamma)\, (1-\sigma)^2}{1+2\gamma} 
\,r^{2\gamma}
\right]\,.
\eea
Hence, if $\gamma\ge2$, the Ricci scalar can be non-singular at the origin, when turning on any
 positive value for $\sigma$. The other curvature invariants built in terms of the Riemann and Ricci tensors show a similar behaviour.

\begin{figure}
\begin{center}
  \includegraphics[width = 0.45 \textwidth]{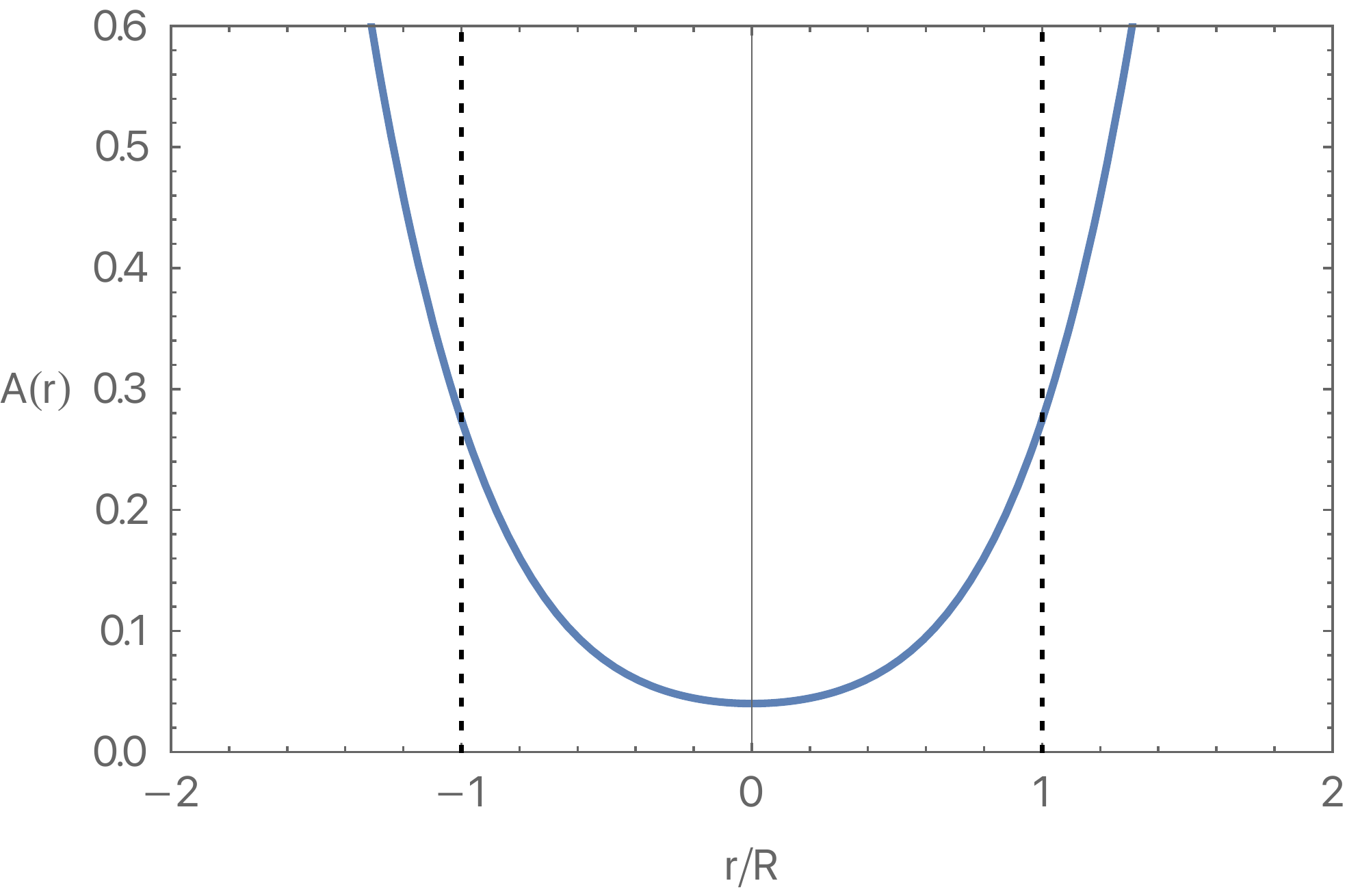}
    \includegraphics[width = 0.45 \textwidth]{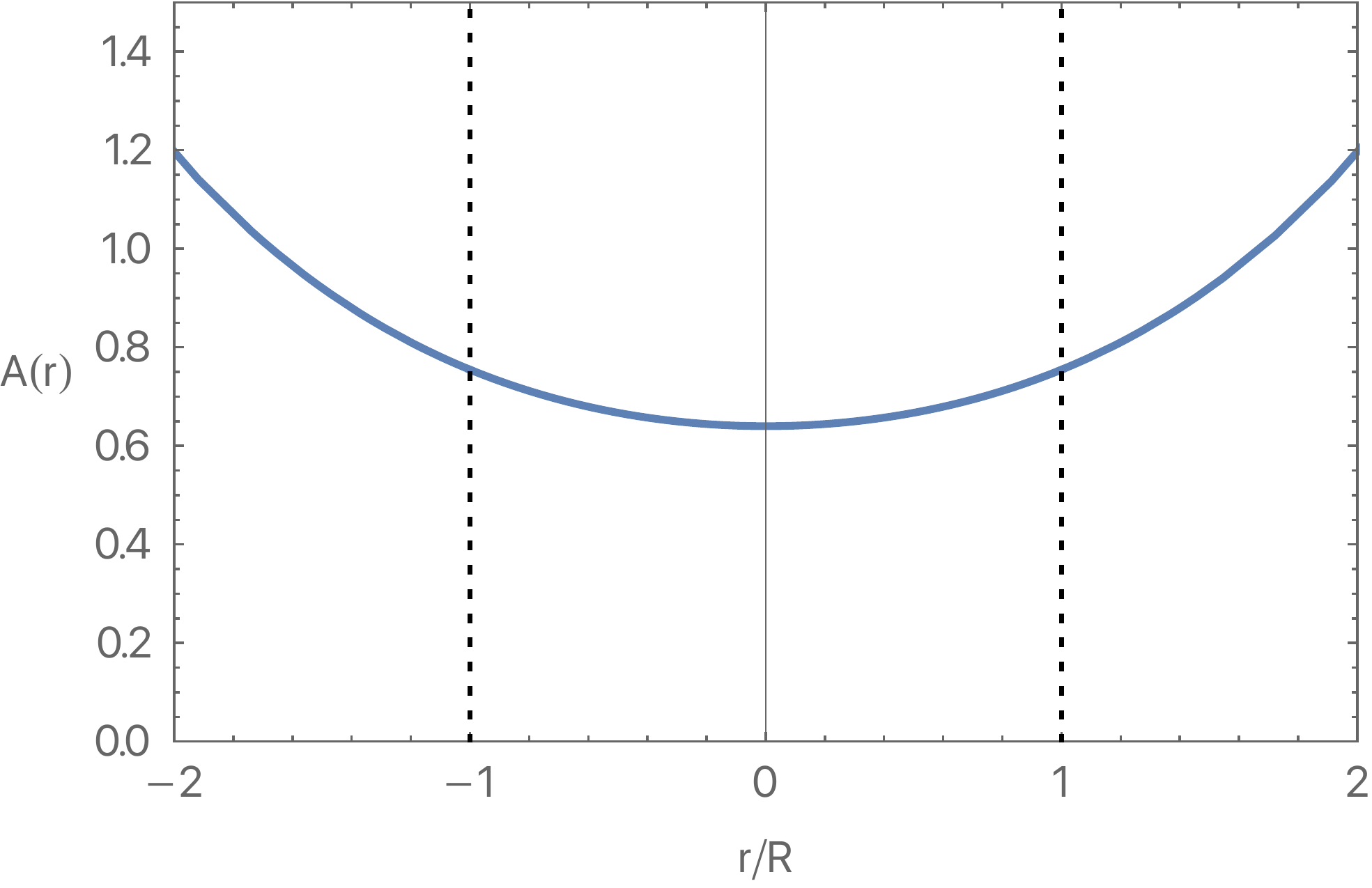}
    \end{center}
 \caption{\small \it The time-time metric profile $A(r)$ of eq \eqref{gensol2a} evaluated nearby the origin,  choosing $\gamma=2$
 and  $\sigma=0.2$ (left panel) and $\sigma=0.8$ (right panel).}
  \label{fig_confinf2}
\end{figure}

We represent in Fig \ref{fig_confinf2} the profile for $A(r)$, which makes manifest that this
function is non-vanishing at the origin and can be continued for negative $r$. One
might explore these configurations as possible wormhole solutions \cite{Morris:1988cz,Visser:1995cc}, and study their properties and their stability. Notice
that the parameter $\sigma$ here is free in the interval $0\,\le\,\sigma\,\le\,1$: it 
would be interesting to determine physical motivations for its size and
how to relate it with observable properties, as for some wormhole 
configurations \cite{Damour:2007ap}.
We postpone these  questions 
to future work.

\section{Geodesics}
\label{sec_geodesics}

We learned in the previous section that -- when considering
spherically symmetric configurations -- the exterior geometry
of our vector
star solutions corresponds to a Schwarzschild space-time. In this
and the next section we discuss possible methods for distinguishing
 our system from GR configurations, and for probing
 the interior properties of the star.

\smallskip
We start studying time-like and null-like geodesics, as a probe of the internal
geometry of the star. 
The behaviour of geodesics around  our  configuration  can be  richer~\footnote{The topic in a similar context has been recently studied in the work \cite{Remmen:2021tyj}.} than
what occurs  in a  Schwarzschild geometry within GR, also thanks to the possibility
of having geodesics crossing the star surface in both directions, probing its interior. 

In fact,
   the analytic study of geodesics
is important for understanding the global and causal properties of the 
geometry under consideration. For  time-like geodesics,
we show that there can be additional stable circular orbits with respect to Schwarzschild black holes.   Geodesics can cross the star surface in both directions, probing
   its 
  compactness. 
 For null-like geodesics, we study the connections between the compactness
of the configuration, and the existence of unstable circular orbits, the light rings associated
with a photosphere \cite{Cardoso:2017cqb}. We assume that the probe particle experiencing the geometry
is uncharged under the vector field $V_\mu$.

We explore the case $\sigma=0$ and focus on the configurations of section \ref{sec_simplest}. Using
standard textbook methods  it is straightforward to obtain the relevant
geodesic equations. We study the trajectory of  a probe particle at radial position   $r(\tau)$, with $\tau$ indicating the particle proper time. We introduce the dimensionless combination
\be
y(\tau)\,\equiv\,\frac{r(\tau)}{R}\,,
\ee
with $R$ indicating the radius of the star. 

\smallskip

For the case of  {\bf
time-like geodesics}, when $y>1$ the particle is outside the star, and experiences
a Schwarzschild space-time characterised by a  mass $M$
 related with $\gamma$ by eq \eqref{relMG1}. The particle geodesics is controlled by the equation
  \be
  \label{tlgeoe1}
\frac12 \left( \frac{d y}{d \tau}\right)^2+U_{\rm ext}(y)\,=\,\frac{\cal E}{R^2}\,\,\,, \hskip1cm {\text{for $y\ge1$.}}
\ee
The geodesic potential is
\bea
U_{\rm ext}(y)&=&\frac{1}{R^2}\,\left[ -\frac{M}{R\,y}
+\frac{\ell^2}{2\,y^2}-\frac{M\,\ell^2}{R\,y^3}
\right]\,,
\nonumber
\\
&=&
\frac{1}{(1+2\gamma)\,R^2}\,\left[ -\frac{\gamma}{y}
+\frac{\ell^2\,(1+2 \gamma)}{2\,y^2}-\frac{\gamma\,\ell^2}{y^3}
\right]\,.
\label{extpotTL}
\eea
In eqs \eqref{tlgeoe1}-\eqref{extpotTL} the quantity ${\cal E}$ is a parameter associated with the conserved energy of the probe particle, and $\ell\,=\,L/R$ is a convenient dimensionless combination, with
 $L$ indicating  the conserved
angular momentum of the particle. 

The particle trajectory in the interior of the star, $y\le1$, is instead described by the equation
 \be
\frac12 \left( \frac{d y}{d \tau}\right)^2+U_{\rm int}(y)\,=\,\frac{\cal E}{R^2}\,\,, \hskip1cm {\text{for $y\le1$.}}
\ee
The  interior potential is
\bea
U_{\rm int}(y)&=&\frac{\left(1+\gamma\right)^2}{2\,R^2\,(1+2\gamma)}\times
\nonumber
\\
&& \hskip-0.3cm
\,\left[
y^{\frac{2 \gamma}{(1+\gamma)}}
+
\left(\frac{ \ell^2}{\left(1+\gamma\right)}+\frac{\gamma^2}{1-\gamma^2}\right)\,
{y^{-\frac{2(1-\gamma)}{1+\gamma}}}
-\frac{1+\gamma\left( 3+\ell^2-\gamma(2+\ell^2)\right)}{\left(1-\gamma^2\right)\,\left(1+\gamma\right)}
 \right]\,.
\label{intpotg1}
\eea

\begin{figure}
\begin{center}
  \includegraphics[width = 0.485 \textwidth]{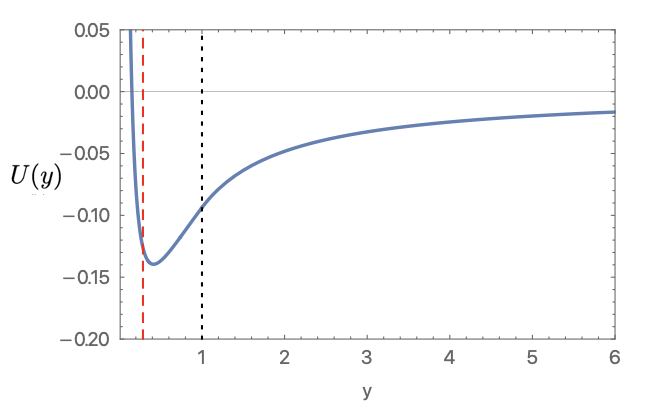}
    \includegraphics[width = 0.485 \textwidth]{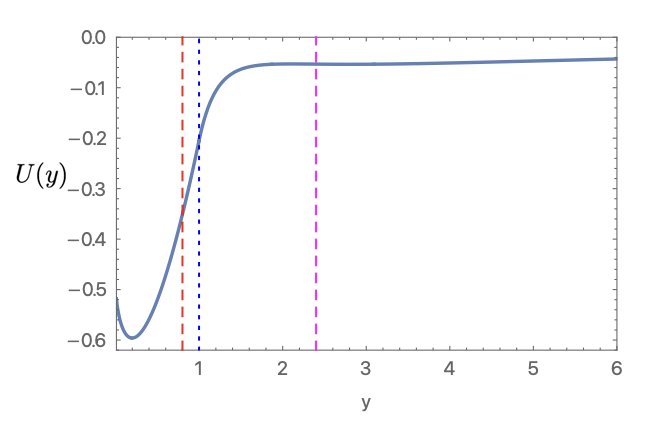}
      \end{center}
 \caption{\small  \it Particle potential for time-like geodesics. Left panel: $\gamma=\ell=1/8$. Right panel $\gamma=2$, $\ell=1.4$. Black dashed line: star surface. Red dashed line: Schwarzschild radius.  On the right we include in magenta dashed line the innermost stable circular orbit position calculated within GR,  and  located at $y\,=\,6\,M\,R$. Both panels
 makes manifest that there are extrema of the geodesic potential located within the star radius. 
 }  \label{fig_timelikegeo}
\end{figure}
 The integration constants characterising  the interior potential \eqref{intpotg1} have been chosen so to ensure that the 
potential $U_{\rm int}(y)$,  together with its first derivative~\footnote{For the special
case $\gamma=1$ it is not possible to ensure continuity of the  first derivative of the potential at the star surface.
},  is continuously connected with  the potential $U_{\rm ext}(y)$  at the star surface $y=1$. When $\gamma<1$ the interior
potential is unbounded and diverges at plus infinity at the origin. 
This implies that a time-like geodesics entering the star
 can not reach the singularity, and bounces back to the exterior (more
 on this later). 
We represent in Fig \ref{fig_timelikegeo} the entire profile of the geodesic time-like potential, for two representative values of $\gamma$.

We can now study more quantitatively the main features of time-like geodesics, using
our expressions \eqref{extpotTL} and \eqref{intpotg1} for the potentials.  In the exterior part of the geometry $y\ge1$ (or 
$r\ge R$), the potential has extrema if the following condition is realised

\be
\ell\,>\,\sqrt{12}\,\frac{\gamma}{1+2\gamma}\,.
\ee
When this inequality is satisfied, the extrema are
 located at the positions
\be
y_{0}\,=\,\frac{\ell^2\,(1+2\gamma)}{2\,\gamma}
\left[
1\,\pm\,\sqrt{1-\frac{12 \gamma^2}{\ell^2(1+2 \gamma)^2}}
\right]\,.
\ee
The plus choice  corresponds a  maximum of the potential (always located  
  in the star
exterior), the minus choice is a minimum of the potential.
 The innermost stable circular orbit
(ISCO) in the star exterior is at 
\be
y_{\rm isco}\,=\,\frac{6 \gamma}{1+2\,\gamma}\,,
\ee
and is outside the star if we choose
 $\gamma>1/4$.

Considering  now the interior geometry, we find that  the potential \eqref{intpotg1} admits an extremal point at
\be
y_0^{\frac{2}{1+\gamma}}\,=\,\frac{\gamma^2+(1-\gamma)\,\ell^2 }{\gamma(1+\gamma)}
\,.
\ee
The position $y_0$ lies within the star interior if the following condition on $\ell$ is satisfied
\bea
\ell^2&<&\frac{\gamma}{1-\gamma} \hskip1cm {\text{for $\gamma<1$}}
\,,
\\
\ell^2&<&\frac{\gamma^2}{\gamma-1}\hskip1cm {\text{for $\gamma>1$}}
\,.
\eea
The value
$\gamma=1$ is special case, since the extremum is always at $y=1/2$ 
(regardless  of the value of $\ell$). When $\gamma>1/4$, by tuning $\ell$, we can have local stable minima for the time-like geodesic
 potential both in the star interior and the star exterior, as shown in Figure \ref{fig_timelikegeo} (right panel).
 
 \smallskip
 It is interesting to study    time-like geodesics which  cross the star surface in both directions: we can use them for probing the star interior.
 In this case, the analysis of the particle trajectories can help in determining the properties
 of the internal geometry.
   The simplest option are radial  plunge orbits, that start at rest from plus infinity with zero angular momentum, and enter
 inside the star radius. Let us then choose ${\cal E}=\ell=0$ in the
 geodesic
 evolution equations. If we further  choose $\gamma<1$, the interior potential (see eq \eqref{intpotg1}) 
 has an infinite barrier at the origin,  which prevents the trajectory from falling into the singularity. A massive  particle arriving from infinitely far away enters into the star region with finite velocity at the star radius, and changes the direction of
 its speed at the zero of the internal potential $U_{\rm int}$. Then
 it bounces, crossing the star surface  
   with the same velocity (in opposite direction) as it enters, and then travels back  to infinite distance from the star. In such a  situation, the propert time $\tau_{\rm int}$ spent within the star in the `bouncing process'  depends only on $\gamma$, and on the star radius. An observer
   far away from the star, who measures the time for radial plunge orbits to reach the star and bounce back, can then infer the star compactness.
   
   The proper time  $\tau_{\rm int}$ spent within the star interior is given by the integral
 \bea
\frac{ \tau_{\rm int}(\gamma, R)}{R}&=&\frac{\sqrt{2}}{R}\,\int_{y_\star}^1\,\frac{d y}{\sqrt{-U_{\rm int} (y)}}
\,,
 \\
 &=&\frac{\sqrt{4+8\gamma}}{1+\gamma}\,\int_{y_\star}^1\,\frac{d y}{y^{\frac{\gamma-1}{\gamma+1}}}\,\left[
 \frac{\left(1+3 \gamma-2 \gamma^2\right)\,y^{\frac{2 (1-\gamma)}{(1+\gamma)}} }{(1-\gamma) (1+\gamma)^2}
 -y^{\frac{2 }{(1+\gamma)}}-\frac{\gamma^2}{1-\gamma^2}
 \right]^{-1/2}\,,
 \label{int2per}
 \eea
 where $y_\star$ is the zero of the potential in the interior -- i.e. the zero of the integrand
 function within square parenthesis in eq \eqref{int2per}. 
 In general,
this expression  needs to be integrated numerically, although for some special values of $\gamma$ an analytical integration is possible. 

For example, for $\gamma=1/2$
we get
\bea
 T_{\rm int}(1/2, R)&=&-\frac{4\,\sqrt{2}}{3}\,\left(1+\frac43\,\arccos\left[-\frac{1}{\sqrt{37}}\right] \right)\,R\,,
 \\
 &\simeq&2.48\,R\,.
 \eea
 Given its dependence on the value of $\gamma$, the time spent by a particle travelling along the geodesics in the interior
 geometry of the star configuration can be used as a tool for studying the star
 properties, as its compactness that depends on $\gamma$ and $R$
 through relation \eqref{defCom1}.

\smallskip
The equation for {\bf null-like} geodesics results

\be
\frac12 y'^2+\frac{1}{R^2}\,\frac{\ell^2}{2\,y^2}\left(1-\frac{2 \gamma}{(1+2\gamma)\,y} \right)\,=\,\frac{{\cal E}^2}{R^2}\,,
\ee
in the exterior. (The various quantities have the same meanings as in the previous time-like case.)
Imposing continuity of the function $y(\tau)$ and its first derivatives at the
star surface, we find the following equation for null-like geodesics  in the  interior
\be
\frac12 y'^2+\frac{1}{R^2}\,\frac{\ell^2}{2\,(1+2\gamma)}\left(
-\gamma+ (1+\gamma)\,y^{-\frac{2 (1-\gamma)}{(1+\gamma)}}
\right)\,=\,\frac{{\cal E}^2}{R^2}\,.
\ee
These results indicate that
 the interior geometry has no extremal points for null-like geodesics. The exterior 
potential admits one 
 maximum 
at
\be
y_{\rm ext}\,=\,\frac{3 \gamma}{1+2\gamma}
\ee
 If $\gamma>1$, $y_{\rm ext}$ 
  is located in the exterior region of the star. Hence,  if this condition is satisfied,
the null-like geodesic potential  have
  a maximum which can be associated with a light-ring and the photosphere: the
 situation is then very similar to GR.  For $\gamma<1$ there is no local maximum instead, and there instead is an infinite barrier for light rays: the situation
 is then similar to what we discussed
 in the case of time-like
 geodesics. The value  $\gamma=1$ is  special because  the potential is constant. The condition for having a photosphere, $\gamma>1$, corresponds
 to the condition of considering  ultracompact objects, ${\cal C}>1/3$, in agreement
 with the general analysis of \cite{Cardoso:2017cqb}.

\section{Beyond spherical  symmetry}
\label{sec_beyond}

We have seen that the behaviour of  geodesics 
can be richer
 with respect to what occurs in a Schwarzschild geometry, since
geodesics can probe the star interior. In
this section we go beyond spherical symmetry, considering the
dynamics of fluctuations of the vector field and of the  metric tensor, both outside
and inside the star. We
will learn that their properties are sensitive to  vector charges
(both electric and magnetic-type ones) as well as the star compactness. They 
can then offer probes for distinguishing a vector star from a black 
hole, as well as to investigate applications of  no-hair 
arguments in this context. 

\smallskip

  We focus on
parity-odd, spin-1 stationary fluctuations: they are easier to deal with analytically,  and  can lead to distinctive effects  
 in   systems (like ours) with a pronounced anisotropic
stress in the  internal stress-tensor of the star.
 Our aims are as follows:
\begin{enumerate}
\item First, in section \ref{vec_susc} we analytically investigate the response of the vector-field profile $V_\mu$
to magnetic-type, spin-1  perturbations, which add  magnetic-type charges to the vector
star solution.
We  determine the structure of the dipolar magnetic field which can be sourced by the star configuration, as well as the vector magnetic susceptibility to the application of an 
external field. 
The characteristic response of the vector 
can be used as a probe
of the star properties, playing a role analog to neutron  star Love numbers \cite{Flanagan:2007ix,Hinderer:2007mb,Damour:2009vw}  for distinguishing exotic compact objects from black holes \cite{Cardoso:2017cfl}. 
\label{pointone}
\item Then, in section \ref{sec_statdef} we focus on metric fluctuations, and study the behaviour of 
the corresponding parity-odd spin-1 deformations. They break spherical symmetry, and can be sourced by magnetic-type
 deformations of the vector profile studied in section \ref{vec_susc}.
 Their properties depend on the electric and magnetic charges,   showing that -- when
breaking spherical symmetry -- the exterior geometry becomes sensitive to the 
 properties of the vector configuration. We are also able to characterize
 the properties of fluctuations in the interior of the star,
    and their dependence on the  vector charge.
 \end{enumerate}

\subsection{Magnetic charge, and the vector magnetic susceptibility}
\label{vec_susc}

We start studying the dynamics of  parity-odd, magnetic  fluctuations of the
vector field profile $V_\mu$, that contribute to  the total combination
\be
\label{vecsplans}
V_\mu \,d x^\mu+\delta V_\mu \,d x^\mu
\,,
\ee
defining the total vector field background$+$fluctuations. We focus on this section on the vector 
fluctuations $\delta V_\mu$ only: we  study their backreaction on the metric in section \ref{sec_statdef}.  
The first part of eq \eqref{vecsplans}, 
 $V_\mu \,d x^\mu$, contains the background electric-type components, as  in    eq \eqref{vecans1}.
 Its profile in the   internal  and external regions  of space-time are given in    
 eqs \eqref{scalALintsol} and \eqref{solVext1} (we focus  on the case of singular
internal geometry of section \ref{sec_simplest}).  
On top, we have
the second contribution to eq \eqref{vecsplans}.  Parity-odd, spin-1 stationary  fluctuations
are parameterised by 
\be
\delta V_\mu \,d x^\mu\,\equiv\, \,\frac{a_{\varphi}(r)}{\sin\theta}\,\left(\partial_\varphi Y_{\ell m}(\theta,\varphi)\right)\,d \theta- \,{a_{\varphi}(r)}\,{\sin\theta}\,\left(\partial_\theta Y_{\ell m}(\theta,\varphi)\right)\,d \varphi\,,
\ee
in terms of a function  $a_{\varphi}(r)$. In this formula the
$Y_{\ell m}$ are the scalar spherical harmonics, and their gradients are associated with 
the spin-1 spherical
harmonics (see e.g. \cite{Maggiore:2007ulw,Hui:2020xxx}) we are interested in. They are
 characterised by a multipole index $\ell$ ($\ell\ge1$) and
an azymuthal index $m$: in our study of fluctuations around spherically symmetric
configurations, there is no dependence on the index $m$ hence we understand it from now on. 
 We assume that the quantities $a_{\varphi}(r)$  are small. We obtain the linearised  equations    in the
 exterior ($r\ge R$) and interior $(r\le R)$ of the star as
 \bea
 \label{extVFL1}
 0&=&
a_{\varphi}''+\frac{2 Q_E\,R}{r (Q_E\,R+ r)}\,a_{\varphi}'-\frac{2 Q_E R+\ell (\ell+1)\,r}{r (Q_E\,R+ r)^2}\,a_{\varphi}\,=\,0\hskip1cm,\,\, %
 \\
 \label{extVFL2}
0&=& a_{\varphi}''+\frac{Q_I (2+\gamma-\gamma^2)-3  \gamma (r/R)^{1+\gamma} }{r \left(Q_I (1+\gamma)+ (r/R)^{1+\gamma}\right)} \,a_{\varphi}'+
\\\hskip-0.1cm
&+&\hskip-0.1cm
\frac{2\,Q_I^2\,\gamma\,
\left( 1+\gamma  \right)
+Q_I   (r/R)^{1+\gamma}
\left( 1+\gamma  \right)
\left( 1+2\gamma  \right)
-
2 (r/R)^{2+2\gamma}
\left( \ell+\gamma (\ell-1) \right)
\left( 1+\ell+ (\ell+2)\gamma \right)
}{2\,
r^2 \left(Q_I (1+\gamma)+ (r/R)^{1+\gamma}\right)^2
}\,a_{\varphi}
\nonumber
\eea
 The internal electric charge $Q_I$ is related with the external one $Q_E$
by the relation \eqref{relQEI}. 
Interestingly, in both the exterior and interior regions the  equations governing the vector
    fluctuations $a_{\varphi}$ are {\it decoupled} from
the parity-odd metric perturbations \footnote{For the case $\ell\ge2$
we need to substitute a constraint condition into the equations -- more 
on this in section \ref{sec_statdef}.}. However, as we will learn, the  vector modes $a_{\varphi}$ backreact on the metric at the linearised level,
breaking spherical symmetry. Notice that also the exterior equation \eqref{extVFL1} depends on the star
electric charge, indicating that the behaviour of the external magnetic deformation  is
controlled by $Q_E$.

The  two equations \eqref{extVFL1} and \eqref{extVFL2} admit two distinct exact solutions, each
depending on two integration constants. Fixing  the integration
constants in the interior so that the solutions continuously match -- together with their
 derivative -- at the star surface $r\,=\,R$, we get
\bea
\label{solaph1}
a_{\varphi}(r)&=&\frac{P_\varphi\,R}{r}\,\left(Q_E+ r/R\right)^{1-\ell} 
+\frac{S_\varphi\,R}{r}\,\left(Q_E+ r/R\right)^{2+\ell} 
\,,
\\
\label{solaph2}
a_{\varphi}(r)&=&\frac{P_\varphi\,R}{r}\,\left(\frac{Q_E (1+\gamma)+\gamma+ (r/R)^{1+\gamma}}{1+\gamma}\right)^{1-\ell} 
\nonumber
\\
&&+\frac{S_\varphi\,R}{r}\,\left(\frac{Q_E (1+\gamma)+\gamma+ (r/R)^{1+\gamma}}{1+\gamma}\right)^{2+\ell} 
\,,
\eea
for two constants $P_\varphi$, $S_\varphi$, respectively in the 
exterior -- eq \eqref{solaph1} -- and in the interior -- eq \eqref{solaph2} -- of
the star configuration.

We now study two  applications of the previous general solutions.
In the first case, we interpret the solutions \eqref{solaph1}-\eqref{solaph2} as controlling
 a  magnetic dipolar potential of the star configuration, with a magnetic
field strength decaying at infinity. Hence, we select $\ell=1$ and $S_\varphi\,=\,0$: the vector potential component $\delta V_\varphi$   becomes
\bea
\delta V_\varphi&=&-
a_{\varphi}(r)\,\sin\theta\, \left(\partial_\theta Y_{10}(\theta)\right)\,,
\nonumber
\\
&=&\sqrt{\frac{3}{4\pi}}\,
P_\varphi\,\frac{R}{r}\,\sin^2\theta\,,
\label{defBrAA}
\eea
for any $r>0$. This  leads  to a dipolar, parity-odd magnetic field deformation $B_{\hat r}$ along the radial direction, which can be computed through the vector field strength using standard formulas
\be
B_{\hat r}\,=\,\frac{\partial_\theta\,V_\varphi}{r^2\,\sin \theta}\,=\,P_\varphi\,
\sqrt{\frac{3}{\pi}}\,
\frac{R}{r^3}\,\cos\theta\,.
\label{defBr}
\ee
The dipolar magnetic field  of the star  is then  controlled by  the 
  magnetic dipolar charge 
  $P_\varphi$. 

In the second case, we study the response of the system
to an external  constant dipolar magnetic field, whose radial component behaves at large distances  ($r\to\infty$) as 
\be \label{BextDEF}
B_{\hat r}^{\rm (ext)}(r\to\infty)\,=\,C_0\,\cos \theta \hskip0.5cm\Rightarrow\hskip0.5cm a_{\varphi}^{\rm (ext)}(r\to\infty)\,=\,C_0\,\sqrt{\frac{\pi}{3}}\,r^2\,,
\ee
for a small  constant $C_0$. In passing from the left to the right of the arrow in
eq \eqref{BextDEF}, we made use of the definitions in eqs \eqref{defBrAA}-\eqref{defBr}. 

 In studying the star vector response to the external field  \eqref{BextDEF} -- a phenomenon that we call vector magnetic  susceptibility -- we impose  for definiteness 
that the magnetic-type fluctuations $a_{\varphi}$ do not blow-up at the origin $r=0$, so to fix 
  the last  of the integration
constants in eq \eqref{solaph2}. The regularity condition requires that $P_\varphi$ and $S_\varphi$ in eq \eqref{solaph2} are
related by the condition
\be
\frac{S_\varphi}{P_\varphi}\,=\,-\frac{\left(1+\gamma\right)^{1+2\ell}}{\left(Q_E(1+\gamma)+\gamma\right)^{1+2\ell}}\,.
\ee
Requiring to match the asymptotic value of the magnetic field as in eq \eqref{BextDEF}, we fix the last of 
the integration constants, and find the following solution for $a_\varphi$:

\bea \label{solresap}
a_\varphi(r)&=&C_0\,\sqrt{\frac{\pi}{3}}\,\frac{R^3}{r}\,\left(Q_E+ r/R
 \right)^{3}
\left[ 1-\left(\frac{R}{r} \right)^{3}
\frac{\left(Q_E+\gamma/(1+\gamma)\right)^{3}}{
\left(Q_E\,R/r+1
 \right)^{3}
}
\,
\right]\,.
\eea
While the part outside the squared parenthesis matches the boundary condition
at spatial infinity, the part inside the parenthesis  controls to the dipolar
susceptibility of the  external magnetic field. Interestingly, such vector magnetic
susceptibility depends both on the electric charge $Q_E$ and the
parameter $\gamma$ controlling the star 
compactness.  This quantity is analogous, although not identical, to the gravitational magnetic
susceptibility as studied in various works in the context of studies of tidal deformability
for black holes
--  see e.g. \cite{Damour:2009va,Kol:2011vg,Landry:2015cva,Hui:2020xxx}.

\begin{figure}
\begin{center}
  \includegraphics[width = 0.45 \textwidth]{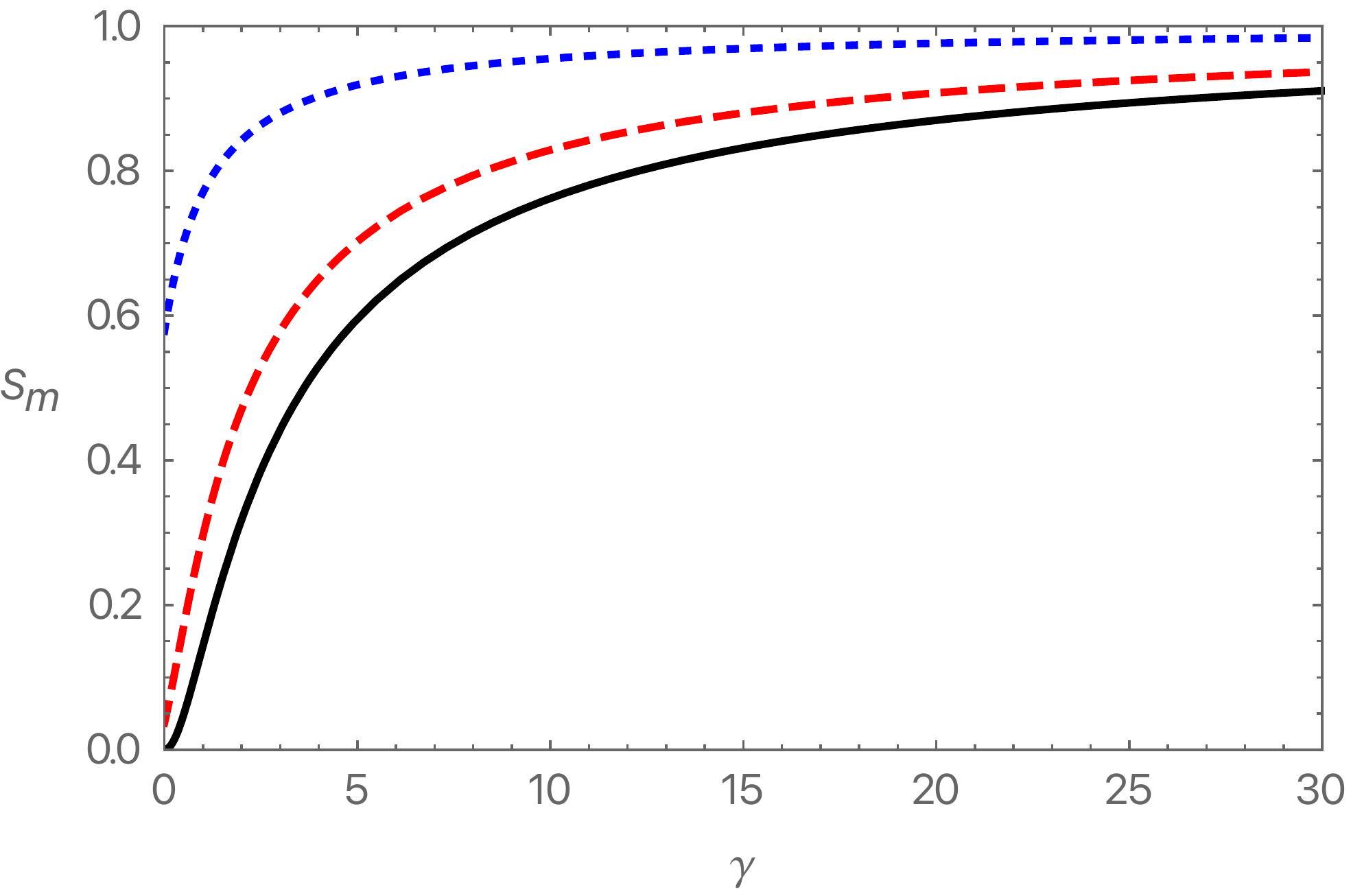}
    \includegraphics[width = 0.45 \textwidth]{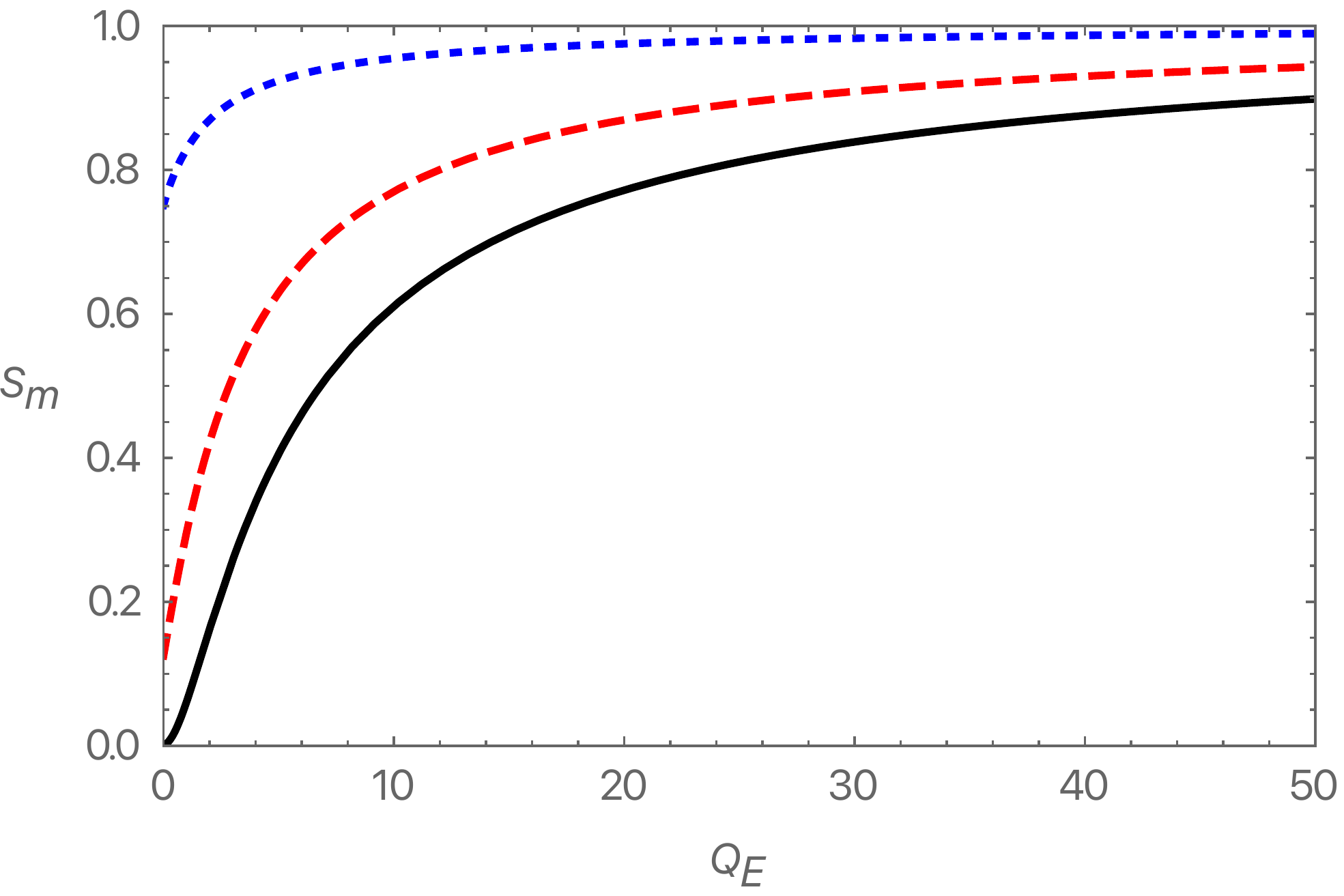}
    \end{center}
 \caption{\small \it Plots of the vector magnetic susceptibility of eq \eqref{defmSM},
 as function of $\gamma$ (left panel) and $Q_E$ (right panel). Left: $Q_E=1/10$ (black); $Q_E=1$ (red dashed); $Q_E=10$ (blue dotted). Right:  $\gamma=1/10$ (black); $\gamma=1$ (red dashed); $\gamma=10$ (blue dotted).  }
  \label{fig_magsus}
\end{figure}

Defining  the vector magnetic susceptibility ${\cal S}_m$ as the opposite of the second
term in the squared parenthesis of eq \eqref{solresap}, evaluated at the star surface $r=R$, we get
\be
\label{defmSM}
{\cal S}_m\equiv 
\frac{\left(Q_E+\gamma/(1+\gamma)\right)^{3}}{
\left(Q_E\,+1
 \right)^{3}}\,.
\ee
Interestingly, this quantity   approaches the value  ${\cal S}_m=1$  for
very large $\gamma$, showing that in the limit of ultracompact objects approaching
the compactness of
a black hole,  the susceptibility behaves as the one of a conducting  sphere.

We plot in Fig \ref{fig_magsus} examples of the dependence of ${\cal S}_m$ on $Q_E$ and
$\gamma$, for some representative cases. It would be interesting to find measurable
observables sensitive to the value of ${\cal S}_m$, which can represent
distinctive probes of the properties of the star -- its charges and compactness.

\subsection{Parity-odd stationary deformations of the metric, and star rotation}
\label{sec_statdef}

After discussing the perturbations in the vector field,
we now focus on parity-odd, stationary fluctuations of the metric,
along the lines of the classical analysis of \cite{Regge:1957td}. 
  To slightly reduce the length of our formulas, in this section we set the star
 radius $R=1$.

 We decompose the metric in a background part and perturbations:
\be
g_{\mu \nu} d x^\mu d x^\nu +h_{\mu\nu}  d x^\mu d x^\nu\,.
\ee
In this equation,  $g_{\mu \nu} d x^\mu d x^\nu$ corresponds to the spherically symmetric background
solution with Ansatz  \eqref{metrans1}, with  exterior and interior metric components studied in section \ref{sec_sphersyms}. The quantity $h_{\mu\nu}  d x^\mu d x^\nu$
controls the parity-odd, spin-1 stationary metric fluctuations in Regge-Wheeler gauge:
\be \small{
h_{\mu\nu}=\begin{pmatrix}
0 & 0& - h_0(r) \sin \theta^{-1}\, \partial_{\varphi} & h_0(r) \sin \theta \,\partial_{\theta} \\
0&0   & -h_1(r)\sin \theta^{-1} \,
  \partial_{\varphi}& h_1(r) \sin \theta \,
   \partial_{\theta} \\
 - h_0(r) \sin \theta^{-1}\,\partial_{\varphi} & -h_1(r)\sin \theta^{-1} \,  \partial_{\varphi} &
   0  & 0 \\
 h_0(r) \sin \theta \, \partial_{\theta} &
   h_1(r) \sin \theta \,\partial_{\theta}& 0 &
  0  \\
\end{pmatrix} } Y(\theta,\varphi).
\ee
The metric fluctuations break the spherical symmetry of the system. 
The angular dependence of the solution is controlled
by derivatives of the scalar spherical harmonics $Y_{\ell  m}(\theta,\varphi)$ ($\ell\ge1$) which define
spin-1 spherical harmonics  as 
in the previous section.     
 It is straightforward
to determine the expression for the linearized equations controlling the radial profiles of the 
metric components $h_0(r)$ and $h_1(r)$, and how they are sourced by 
the magnetic-type vector fluctuations studied in section \ref{vec_susc}. 
 For the case $\ell=1$, the function $h_1(r)$
is a  gauge mode and can be set to zero; for $\ell\ge2$, the equation of $h_1$ is algebraic, and can be solved as a function of $h_0$ and $a_\varphi$, both in the exterior and the interior of the star. For example, in the star exterior  we find
\be
h_1(r)\,=\,\frac{r}{r-2M}\,\frac{\sqrt{Q_E^2+2 Mr + Q_E r}}{Q_E+r}\,h_0(r)-\frac{r\,\sqrt{Q_E^2+2\, M\, r+Q_E \,r}}{2 (Q_E+r)^2}\,a_\varphi(r)\,.
\ee
Plugging this expression into the equation for $a_\varphi$ we find equation \eqref{extVFL1}, that
-- as explained in section \ref{vec_susc} -- is decoupled from metric fluctuations. The quantity $h_0(r)$
is controlled by the following equation in the star exterior, valid for any $\ell\ge1$ (we
understand the dependence on $r$):
\bea
0&=&h_0''-\frac{4 Q_E}{r\,(Q_E+r)}\,h_0'+\frac{\left( 6 Q_E^2+4 \,Q_E\,r- \ell (\ell+1) r^2\right)}{r^2 \,\left( Q_E+r\right)^2}\,h_0
+\frac{4 M-2 r}{Q_E+r}\,a_\varphi''-\frac{4 M}{r}\,\frac{a_\varphi'}{Q_E+r}
\nonumber
\\
&+&
\frac{a_\varphi}{r\,(Q_E+r)^3}
\left[ Q_E^2 \left(1+\frac32 \ell (\ell+1) \right) +2 M r+\left(1+ \frac32 \ell (\ell+1)\right) Q_E r- \ell (\ell+1) (M-2 r)\,r\right]\,.
\nonumber
\\
\label{eqh0stext}
\eea
Notice that the metric component $h_0(r)$ is sourced
by the magnetic-type vector perturbation $a_\varphi(r)$. The equation
for $h_0$ in the star interior can also be derived quite easily, but is longer and we
do not write it explicitly.

\smallskip
Let us study explicit examples of solutions. Focus on the dipolar case $\ell=1$, $m=0$ (and $h_1=0$), and  on perturbations that decay at infinity. As first noticed in \cite{Regge:1957td},  the metric then  describes a slowly rotating  configuration, with $h_0(r)$ controlling the   stationary rotation
of the geometry
through an off-diagonal metric component
\be
g_{t\varphi}\,dt \,d \varphi\,=\,-2\,\left(\sqrt{\frac{3}{ 4\pi}}\, \frac{h_0(r)}{r^2}\right)\,r^2\,\sin^2 \theta \,dt \,d \varphi\,.
\ee
 It
is straightforward to solve  eq \eqref{eqh0stext} for $h_0(r)$ in the star exterior, taking as source the magnetic-type solution we found in eq \eqref{solaph1} with $\ell=1$ and $S_\varphi=0$. We find for $r\ge1$

   \be
   \label{eqh0solE1}
-\sqrt{\frac{3}{ 4 \pi}}\,\frac{h_0(r)}{r^2}\,=\,\frac{a\,M}{(Q_E+r)^3}+\sqrt{\frac{3}{ \pi}}\,\frac{P_\varphi\,M}{Q_E\,r^3}\,,
\ee
with $a$ an integration constant associated with the rotation of the geometry (the Kerr parameter), and $P_\varphi$ the intrinsic
magnetic dipolar charge as given in eq \eqref{defBrAA} above.  This metric
coefficient should be compared with the one
 found in the slowly-rotating limit of the Kerr solution in GR:
\be
   \label{eqh0solE1GR}
-\sqrt{\frac{3}{ 4 \pi}}\,\frac{h^{\rm GR}_0(r)}{r^2}\,=\,\frac{a\,M}{r^3}
 \,.
\ee
Comparing \eqref{eqh0solE1} with \eqref{eqh0solE1GR}, we find that the magnetic charges $P_\varphi$
and $Q_E$ influence the geometry as vector hairs, and  contribute in breaking the 
spherical symmetry of the configuration to an axial one. Interestingly, $Q_E$ appears at the denominators: we interpret this fact as a consequence of  the specific non-minimal
couplings of the vector to gravity. The quantity  $Q_E$ contributes to modulate
 the $1/r^3$ decay at large distances of the angular momentum contribution proportional
 to $a$. Instead, $P_\varphi/Q_E$ adds a new  contribution to $h_0$, absent in the vacuum GR case, being induced by the magnetic dipolar charge of the star configuration.
 
 The peculiar dependence on the radial coordinate $r$ in eq \eqref{eqh0solE1} can
 lead to interesting effects, distinctive of the vector-tensor system under consideration. Let 
 us suppose that $a$ and $P_\varphi$ have opposite sign (say $a$ is positive) and call 
 \be
 \nu^3\,\equiv\,-\sqrt{{3}/{ \pi}}\,{|P_\varphi |\,M}/{(Q_E\,a)}
 \ee 
 We assume $0\,\le\,\nu\,<\,1$.
We can rewrite eq  \eqref{eqh0solE1} as
\be
\label{eqh0solE1a}
-\sqrt{\frac{3}{ 4 \pi}}\,\frac{h_0(r)}{r^2}\,=\,\frac{a\,M}{r^3}
\left(\frac{r^3}{(Q_E+r)^3}- {\nu^3}
\right)\,.
\ee
While for large values of $r$ the right hand side of \eqref{eqh0solE1a} is positive, as we approach the star the sign of $h_0(r)$ changes. This occurs at
\be
r_\star\,=\,\frac{\nu\,Q_E}{1-\nu}
\ee
which is  larger than the star radius $R=1$ by choosing a large enough $Q_E$. This fact implies that the direction of rotation of the space-time changes as we approach the star, due to to the opposite contributions to rotation of the Kerr parameter
$a$ and the magnetic charge $P_\varphi$. It would be interesting to further explore the consequences of this phenomenon, and whether can be used to find distinctive probes
of this system. 

\smallskip

The discussion of the interior part of the geometry is particularly simple in the case $P_\varphi\,=\,0$: let us focus for simplicity on this case for the rest of the section, although analytic expressions for $P_\varphi\,\neq\,0$ can also be found. This limit implies that the exterior dipolar configuration is described by eq \eqref{eqh0solE1} with no magnetic-charge contribution. The interior solution $r\le1$ of the field equations for $h_0$ that continuously match
with the exterior for  $r\ge1$ is
\bea
-\sqrt{\frac{3}{ 4 \pi}}\,\frac{h_0(r)}{r^2}&=&\frac{a \,M \,(1+\gamma)^3}{\left[Q_E(1+\gamma)+\gamma (1-\sigma)+  
\,r^{1+\gamma} (1-\sigma)+\,r\, \sigma \,(1+\gamma)\right]^3}\,,
\eea
where we selected as interior background configuration the solution described in section \ref{sec_resolving}. The interior part of the star is dragged by the rotation of the configuration: the dragging effects are also modulated by the electric charge $Q_E$
and depend on the star compactness through the parameter $\gamma$. 
While in this example we focussed on the case $\ell=1$, the cases $\ell\ge2$ can
also be straightforwardly analysed, at least in absence of intrinsic magnetic charge, and the previous formulas  generalise by changing the exponents from $3$ to $\ell+2$.  

\smallskip

In summary, the study of properties of parity-odd stationary metric fluctuations reveal 
a very rich structure that goes beyond what found in solutions of GR in empty space, even
if the exterior star geometry is given by Schwarzschild space-time. Both the
 exterior and the interior metric elements are  indeed sensitive to the vector properties. A  detailed
analysis of time-dependent perturbations, as well as parity-even modes,  is left for future studies.

\section{Outlook}
\label{sec_outlook}

In this work we presented
and investigated  new analytic solutions
 describing a family of  horizonless compact objects in vector-tensor
 theories of gravity, dubbed ultracompact vector stars.  They are sourced by
  a vector condensate,  induced by a non-minimal coupling with gravity. They
  can 
  be as compact as black
 holes, thanks to their internal anisotropic stress. In the spherically symmetric case, their
 external geometry corresponds to the Schwarzschild solution as
 in GR, even if they can be characterised by an electric-type charge. 
   The interior part of the solution  resembles
   a  singular isothermal sphere, and the singularity at the star centre can be resolved by
    tuning some of the available integration constants. The interior configuration  
    is smoothly matched to the exterior geometry of the star, with no need of extra stress-tensor 
    on the star surface.  
 
 We investigated features of our systems that  allow one to distinguish vector star objects from GR black hole solutions. 
 We analytically studied the 
   behaviour of geodesics trajectories within the star interior -- where new stable circular orbits are allowed -- as well as geodesics crossing in both ways the star surface. 
  We analytically investigated parity-odd stationary fluctuations that break spherical
  symmetry and can assign a magnetic charge to our vector star configurations.

   It would be  interesting to further develop our analysis  along the  following directions:
   \begin{itemize}
   \item Study time-dependent fluctuations -- both odd and even parity -- and investigate  the stability 
the horizonless vector star objects. This topic is particularly interesting since black hole
configurations in vector-tensor  theories of gravity have been shown to have instabilities \cite{Kase:2018voo,Garcia-Saenz:2021uyv} which might be cured (or not) in more general configurations like ours.
\item Study mechanisms  of formation of vector stars  from a process of  gravitational collapse,   including further  sources of energy momentum tensor in the form
 of standard matter.   Gravitationally bound
   solitons made of  dark-matter vector bosons -- in a Proca theory  -- have been recently numerically investigated in \cite{Adshead:2021kvl,Jain:2021pnk,Gorghetto:2022sue,Amin:2022pzv}, and would be
   interesting to pursue similar studies in our context. In fact, non-minimal couplings
   of the vector with gravity are allowed (and expected) once the vector 
   Abelian symmetry is broken. 
   \item Determine solutions with a large rotation parameter in the exterior geometry -- the analog of Kerr -- and large magnetic charge. Investigate the properties and phenomenology 
   of the resulting magnetically-charged  configurations, for example taking
   inspiration from \cite{Maldacena:2020skw}. 
 \item Determine additional field-theory and cosmological motivations for the vector-tensor theories we consider and their possible extensions for  dark matter
 and dark energy. 
   \end{itemize}    
We leave these points to future investigations.

\section*{Acknowledgements}
It is a pleasure to thank Mustafa Amin and Ivonne Zavala for discussions.
The work is funded by the STFC grant ST/T000813/1. For the purpose of open access, the author has applied a Creative Commons Attribution  licence to any Author Accepted Manuscript version arising.

{\small
\providecommand{\href}[2]{#2}\begingroup\raggedright\endgroup
}

\end{document}